%% file: 24366.tex
\documentclass[dvipsnames]{aa}
\pdfoutput=1

\ifx\pdfoutput\undefined
\usepackage{graphicx}
\else
\usepackage[pdftex,pdfauthor={J.~Threlfall},citecolor=blue,linkcolor=blue, linktocpage=true,colorlinks=true]{hyperref}
\usepackage{epstopdf}
\fi 

\usepackage{natbib}
\bibpunct{(}{)}{;}{a}{}{,} 
\usepackage{txfonts}
\usepackage[usenames]{color}
\usepackage{subfig}	
\usepackage{epsfig}	
\usepackage{cancel}
\usepackage{multirow}
\usepackage{fixltx2e}	
\usepackage{amssymb}	
\graphicspath{{figs/}}

\newsavebox{\bigleftbox}

\input{math_def.tex}	
\date{Received 10 June 2014 / Accepted 27 October 2014}			

\begin{document}
 \title{Particle acceleration at a reconnecting magnetic separator}
 \titlerunning{Particle acceleration at a reconnecting magnetic separator}
 \author{J.~Threlfall \and T.~Neukirch \and C.~E.~Parnell \and S.~Eradat~Oskoui}
 \institute{School of Mathematics and Statistics, University of St Andrews, St Andrews, Fife, KY16 9SS, U.K. \email{\{jwt9;tn3;cep;se11\}@st-andrews.ac.uk}\label{inst1}}
 \abstract
 {While the exact acceleration mechanism of energetic particles during solar flares is (as yet) unknown, magnetic reconnection plays a key role both in the release of stored magnetic energy of the solar corona and the magnetic restructuring during a flare. Recent work has shown that special field lines, called separators, are common sites of reconnection in 3D numerical experiments. To date, 3D separator reconnection sites have received little attention as particle accelerators.}
 {We investigate the effectiveness of separator reconnection as a particle acceleration mechanism for electrons and protons.}
 {We study the particle acceleration using a relativistic guiding-centre particle code in a time-dependent kinematic model of magnetic reconnection at a separator.} 
 {The effect upon particle behaviour of initial position, pitch angle, and initial kinetic energy are examined in detail, both for specific (single) particle examples and for large distributions of initial conditions. The separator reconnection model contains several free parameters, and we study the effect of changing these parameters upon particle acceleration, in particular in view of the final particle energy ranges that agree with observed energy spectra.}
 {}
 \maketitle

\section{Introduction}\label{sec:Intro}

Understanding the physical processes causing the acceleration of a large number of charged particles to high energies during solar flares is one of the biggest unsolved problems in solar physics \citep[e.g.][]{review:Fletcheretal2011}. One of the mechanisms that is strongly linked with particle acceleration in solar flares is magnetic reconnection \citep[see e.g.][]{paper:Neukirchetal2007}.

Magnetic reconnection is a fundamental mechanism that lies at the heart of many dynamic solar (stellar), magnetospheric and astrophysical phenomena. It is required to enable local and global restructuring of complex magnetic fields; in so doing, it (crucially) allows stored magnetic energy to be released in the form of bulk fluid motion (waves) and/or thermal/non-thermal energy (i.e. local heating and/or high energy particles). In this context, particle acceleration mechanisms have been widely studied, mainly for two reasons: (a) There is a general consensus that magnetic reconnection plays a major role in the release of magnetic energy and its conversion into other forms of energy during flares. (b) Magnetic reconnection is generically associated with parallel electric fields \citep[e.g.][]{paper:Schindleretal1988,paper:HesseSchindler1988,paper:Schindleretal1991}, hence should lead to particle acceleration. Actually, the concept of magnetic reconnection was first introduced in order to explain the possible generation of high energy particles in flares \citep{paper:Giovanelli1946}.

As a consequence of the vast difference in length and time scales between the macroscopic (MHD or magnetohydrodynamic) description of reconnection in solar flares and the microscopic description of particle acceleration, most studies use a test particle approach. A large proportion of these past studies of particle acceleration by magnetic reconnection have focussed on acceleration in 2D or 2.5D reconnection models. Typically, these are either (two-dimensional) null point configurations or current sheets, or a combination of both, in many cases including a guide field in the invariant direction \citep[e.g.][]{paper:BulanovSasorov1976,paper:BruhwilerZweibel1992,paper:Kliem1994,paper:Litvinenko1996,paper:BrowningVekstein2001,paper:ZharkovaGordovskyy2004,paper:ZharkovaGordovskyy2005,paper:WoodNeukirch2005,paper:HannahFletcher2006,paper:Drakeetal2006,paper:Gordovskyyetal2010a,paper:Gordovskyyetal2010b}.

Over the past decade there has been an increasing number of studies of particle acceleration in 3D reconnecting magnetic field configurations. This includes, for example, test particle calculations at 3D magnetic null points (e.g. \citealp[][]{paper:DallaBrowning2005,paper:DallaBrowning2006,paper:DallaBrowning2008,paper:Guoetal2010,paper:Stanieretal2012}, and more recently also PIC simulations, see e.g. \citealp[][]{paper:Baumannetal2013}) and in magnetic configurations undergoing magnetic reconnection at multiple sites \citep[e.g.][]{paper:Vlahosetal2004,paper:ArznerVlahos2004,paper:ArznerVlahos2006,paper:Turkmanietal2005,paper:Turkmanietal2006,paper:Brownetal2009,paper:GordovskyyBrowning2011,review:Cargilletal2012,paper:Gordovskyyetal2013,paper:Gordovskyyetal2014}.

In this paper, we will investigate particle acceleration in a magnetic reconnection configuration which has so far not received any attention in relation to particle acceleration, namely a reconnecting 3D magnetic separator. Separators are special magnetic field lines which join pairs of magnetic null points and lie at the intersection of four distinct flux domains. Although the magnetic configurations about separators can be loosely regarded as the 3D analogue of 2D X-point (and O-point) plus guide field configurations, which have been studied in connection with particle acceleration before, recent advances in theory and computational experiments have shown that reconnection of 3D magnetic field configurations is fundamentally different to the widely used 2D (or 2.5D) reconnection models. 

In 2D models, the reduced degree of freedom requires that reconnection only takes place at an X-type null point (where magnetic flux is carried in towards the null where field-lines are cut and rejoined in a one-to-one pairwise fashion before being carried away). However, in 3D, the presence of a localised non-ideal region (where there exists some component of electric field parallel to the magnetic field) means that this simple "cut and paste" picture of field line reconnection no longer holds; instead magnetic flux is reconnected continually and continuously throughout the non-ideal region \citep[for reviews of 3D magnetic reconnection, see e.g.][]{book:PriestForbes, book:BirnPriest,review:Pontin2011}. Hence, one could view the work presented here as a natural extension of previous work to three dimensions. However, the role that reconnecting magnetic separators play in the acceleration of particles is, as yet, unknown.

Despite being defined as magnetic field lines connecting two magnetic null points, fundamentally, separator reconnection is an example of non-null reconnection \citep{paper:Schindleretal1988,paper:HesseSchindler1988}. It is known that separators are prone to current sheet formation \citep{paper:LauFinn1990, paper:Parnelletal2010b,subm:Stevensonetal2014} and thus are likely sites of magnetic reconnection. Theoretical studies of separator reconnection have continued to evolve over many years \citep[e.g.][]{paper:Sonnerup1979,paper:LauFinn1990,paper:LongcopeCowley1996,paper:Galgaardetal2000,paper:Longcope2001,paper:PontinCraig2006,paper:Parnelletal2008,paper:DorelliBhattacharjee2008}. Building on experiments by \citet{paper:Haynesetal2007} and \citet{paper:Parnell2007}, recent work has highlighted that multiple magnetic separators may exist within a given magnetic environment at any given time \citep{paper:Parnelletal2010a} and that separator reconnection is an important and fundamental process when emerging magnetic flux interacts with overlying magnetic field \citep{paper:Parnelletal2010b}. More recently, \citet{paper:Wilmot-SmithHornig2011} have shown that even simple magnetic configurations may evolve into configurations containing multiple separators. 

Observationally, separator reconnection has been inferred via EUV observations of the solar corona \citep{paper:Longcopeetal2005}, while in situ measurements from {{Cluster}} have also identified reconnecting magnetic separators in Earth's magnetosphere \citep[e.g.][]{paper:Xiaoetal2007,paper:Dengetal2009,paper:Guoetal2013}. While there are no direct observations of any topological features such as null points or separators during a flare, evidence from magnetic field models extrapolated from magnetograms suggest that separators are indeed likely reconnection sites within flares. In addition, there is growing observational support for particle acceleration models which energise particles through magnetic reconnection at separators \citep[][]{paper:Metcalfetal2003}; a broad overview of this subject may be found in \citep{review:Fletcheretal2011}.

The primary objective of the present work is to determine what role (if any) separators may play in the acceleration of particles, in a general context rather than in the particular case of a flare. We specifically investigate how test particle orbits and energy gains depend on initial conditions and how observations (for example, of solar flares) may be used to constrain our separator reconnection model parameters.
In Sec.~\ref{sec:model} we discuss the model itself, comprising of a global field \citep[based on the kinematic model of][described in Sec.~\ref{subsec:globalfield}]{paper:Wilmot-SmithHornig2011} into which we place test particles (whose governing equations are outlined in Sec.~\ref{subsec:guidingcentre}). We investigate the role of several initial conditions in the recovered particle behaviour in Sec.~\ref{sec:localbehaviour}, before studying larger distributions of particles in Sec.~\ref{sec:globalbehaviour}. A discussion of our findings is presented in Sec.~\ref{sec:discussion} before conclusions and future areas of study are outlined in Sec.~\ref{sec:conclusions}.

\section{Model setup}\label{sec:model}
Our model can be broadly split into two parts: a (time-dependent) large-scale electromagnetic field environment, into which we insert particles and the test particle motion itself, which is modelled using the relativistic guiding centre approximation. Details of these two parts are described in the following two subsections.

\subsection{Global field}\label{subsec:globalfield}
We base our global separator field model on that of \citet{paper:Wilmot-SmithHornig2011}. The initial magnetic field is a potential one, of the form 
\begin{equation}
 {\bf{B}}_0=\frac{b_0}{L^2}\left[x(z-3z_0){\xhat}+y(z+3z_0){\yhat}+\frac{1}{2}\left( z_0^2-z^2+x^2+y^2\right) {\zhat} \right],  \\
 \label{eq:B0}
\end{equation}
with magnetic null points at $(0,0,\pm{z_0})$; $b_0$ and $L$ determine the characteristic field strength and length scale of the model. For the original (essentially scale-free) model of \citet{paper:Wilmot-SmithHornig2011}, $z_0=5$, $b_0=1$ and $L=1$.

\begin{figure}[t]
 \centering
 \resizebox{0.45\textwidth}{!}{\includegraphics{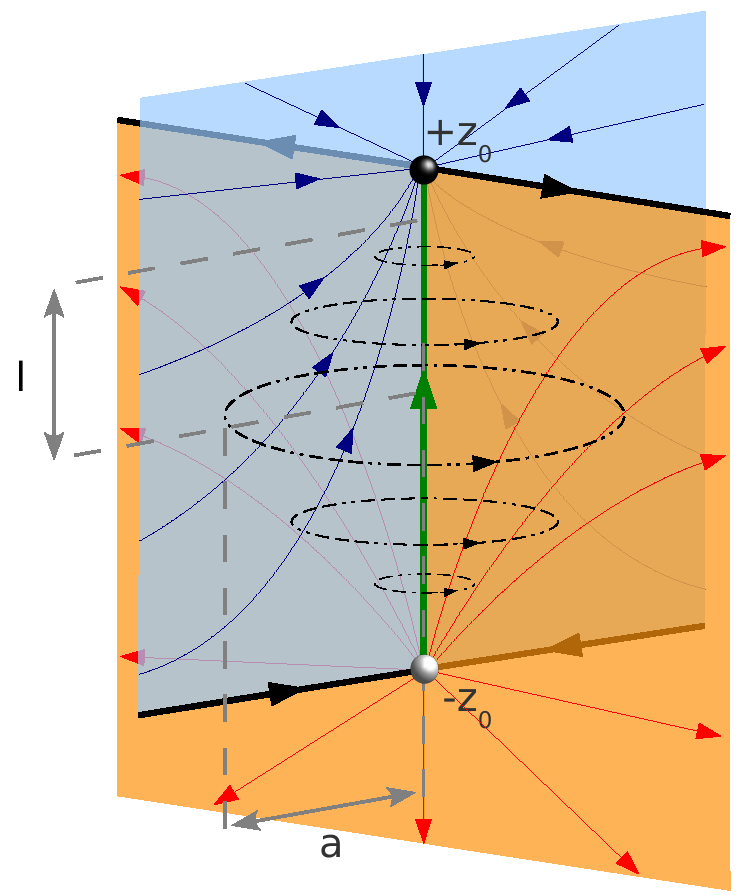}} 
 \caption{Cartoon illustrating the separator reconnection model of \citet{paper:Wilmot-SmithHornig2011}. A black (white) sphere indicates the location of the upper (lower) null, the fan plane of which is seen in blue (orange) while the spines of both nulls are shown as thick black lines (due to the fan plane transparency, these lines adopt the colour of any fan planes they pass behind). Field lines are also included on each fan plane to indicate local magnetic-field orientation, coloured to match a specific plane. A separator (shown in green) links both nulls at the fan plane intersection. Time-dependent rings of magnetic flux (dot-dashed rings) induce an electric field directly along the separator (and the near vicinity); the horizontal and vertical extent of this perturbation are controlled by parameters $a$ and $l$ in Eq.~\ref{eq:Br}.} 
 \label{fig:WS}
\end{figure}
The null at $(0,0,-z_0)$ is classified as a \emph{positive null}, while that at $(0,0,+z_0)$ is a \emph{negative null} (due to the orientation of magnetic field in the spine/fan at each null). The magnetic separator is formed by the intersection of the two fan planes associated with each null. The separator and general configuration of this model can be seen in Fig.~\ref{fig:WS}. 

With Faraday's law coupling (time-varying) magnetic and electric fields, introducing a ring of magnetic flux of the form
\begin{equation}
 {\bf{B}}_{r}=\curl{\left[b_1 a \exp{\left(-\frac{\left( x-x_c\right)^2}{a^2}-\frac{\left( y-y_c\right)^2}{a^2}-\frac{\left( z-z_c\right)^2}{l^2} \right){\zhat} } \right] },
 \label{eq:Br}
\end{equation}
(centred on $(x_c,y_c,z_c)$, with radius (in the $xy$ plane) controlled by the parameter $a$, the height (in $z$) by $l$ and the field strength by $b_1$) induces a vertical electric field with the form
\begin{equation}
 {\bf{E}}=-\frac{b_1 a}{\tau}\exp{\left(-\frac{(x-x_c)^2}{a^2}-\frac{(y-y_c)^2}{a^2}-\frac{(z-z_c)^2}{l^2} \right) } {\zhat}.
 \label{eq:E}
\end{equation}
(provided that the time evolution satisfies Faraday's law, i.e. that
\begin{equation}
 {\bf{B}}={\bf{B}}_0+\frac{t}{\tau}{\bf{B}}_{r}, \qquad 0\leq t\leq\tau,
 \label{eq:B}
\end{equation}
taking place over a timescale $\tau$). By setting $x_c~=~y_c~=~z_c~=~0$, an anti-parallel electric field is induced along the separator (and in the local vicinity). 

By taking the curl of Eq.~\ref{eq:B}, it can be shown that the time evolution of current broadly agrees with numerical 3D simulation models of magnetic separator reconnection, where parallel electric currents (and hence parallel electric fields) are typically seen to accumulate about the magnetic separator \citep[see e.g.][]{paper:Parnelletal2010a,paper:Parnelletal2010b}. Furthermore, while the original forms of magnetic and electric fields are constructed by applying Faraday's Law, they also satisfy a generalised Ohm's Law \citep[see discussion in][for more details]{paper:Wilmot-SmithHornig2011}.

In order to make these equations dimensionless, we define the dimensions of our model through a field strength $\bscl$, lengthscale $\lscl$ and timescale $\tscl$; dimensional and dimensionless quantities are related via
\[
{\bf{B}}=\bscl\,\bar{\bf{B}}, \qquad x=\lscl\,\bar{x}, \qquad t=\tscl\,\bar{t},
\]
where barred quantities represent dimensionless counterparts of the relevant variables. These quantities also fix other normalising constants; for example velocities in the model are scaled by $\vscl(={\lscl}{\tscl}^{-1})$,  energies by $\KEscl(=0.5m{\vscl}^2)$ and (assessing the dimensions of Faraday's Law) electric fields are scaled by $\escl(={\bscl\,\lscl}{\tscl}^{-1}=\bscl\,\vscl)$.

This investigation is motivated by situations which might be found in the solar atmosphere. We therefore fix our normalising quantities appropriately; in this experiment we take $\bscl=0.01\rm{T}$, $\lscl=10\rm{Mm}$ and $\tscl=100\rm{s}$, and select $\tau=100\rm{s}$, $b_0=0.01\rm{T}$ and $b_1=20b_0$ for simplicity. 

At present, we are unable to estimate the size of a typical width of a reconnection region/current sheet purely through observations; kinetic studies of magnetic reconnection suggest that a current sheet width which approaches 10 ion inertial lengths ($10c/\omega_{pi}$, where $\omega_{pi}$ is the local ion plasma frequency) may not be unrealistic \citep[see e.g.][and references therein]{paper:WoodNeukirch2005}. Assuming a typical coronal number density of $10^{15}{\rm{m}}^{-3}$, we find that $10c/\omega_{pi}\simeq72\rm{m}$ for singly ionised hydrogen. We will use this value to constrain the selection of parameters $a$ and~$l$. By comparison, assuming a coronal temperature of $2\times10^6\rm{K}$, typical electron/ion gyroradii under the same conditions would be $0.31\rm{cm}$/$13.4\rm{cm}$ respectively.

\subsection{Relativistic particle dynamics}\label{subsec:guidingcentre}
Having now established the global environment into which these particles will be inserted, we briefly turn our attention to the details of the particle motion itself. In anticipation of particle velocities which are a significant fraction of the speed of light ($c$), we utilise the full relativistic set of guiding-centre-motion equations, outlined in \citet{book:Northrop1963} \citep[based on the treatment of][]{paper:Vandervoort1960}, presented here in normalised form:
\begin{subequations}
 \begin{align}
  \frac{d{u_\parallel}}{dt}&=\frac{d}{dt}\left(\gamma\vpar\right)=\gamma\ue\cdot{\frac{d{\bf{b}}}{dt}}+\omscl\tscl E_\parallel-\frac{\mu_r}{\gamma}\frac{\partial{B}}{\partial s}, \label{eq:Rnorm1} \\
  {\bf\dot{R}_\perp}&=\ue+\frac{\bf{b}}{B^{\star\star}}\times\left\lbrace \frac{{1}}{\omscl\tscl}\left[ \frac{\mu_r}{\gamma}\left( \grad{B^\star}+ \frac{\vsclsq}{{c^2}}\ue\frac{\partial B^\star}{\partial t}\right)\right.\right. \nonumber \\ 
   &\qquad\quad\qquad\qquad\left.\left. +u_\parallel\frac{d{\bf{b}}}{dt}+\gamma\frac{d\ue}{dt}\right]+\frac{\vsclsq}{{c^2}}\frac{u_\parallel}{\gamma}{E_\parallel}\ue \right\rbrace, \label{eq:Rnorm2} \\ 
  \frac{d\gamma}{dt}&=\frac{\vsclsq}{{c^2}}\left[\omscl\tscl\left({\bf\dot{R}_\perp}+\frac{u_\parallel}{\gamma}{\bf{b}}\right)\cdot{\bf{E}}+\frac{\mu_r}{\gamma}\frac{\partial B^\star}{\partial t}\right],   \label{eq:Rnorm3} \\
  \mu_r&=\frac{\gamma^2{\vperp^2}}{B}. \label{eq:Rnorm4}  
 \end{align}
 \label{eq:rel_norm} 
\end{subequations}
{\noindent}Here $\mu_r$ is the relativistic magnetic moment (often expressed in terms of momentum $p_\perp$), for a particle with rest-mass $m_0$ and charge $q$, whose guiding centre is located at ${\bf{R}}$, subject to a magnetic field ${\bf{B}}$ (with magnitude $B(=|{\bf{B}}|)$ with a unit vector ${\bf{b}}(={\bf{B}}/B)$) and an electric field ${\bf{E}}$. Depending on the local conditions, this particle is likely to experience guiding centre drifts; the largest in magnitude is typically the ${E}\times{B}$ drift, which has a velocity $\ue(={\bf{E}}\times{\bf{b}}/B)$. The component of velocity parallel to the magnetic field is $\vpar(={\bf{b}}\cdot{\dot{\bf{R}}})$, while $\Epar(={\bf{b}}\cdot{\bf{E}})$ is the magnitude of the electric field parallel to the local magnetic field, $\dot{\bf{R}}_\perp(=\dot{\bf{R}}-\vpar{\bf{b}})$ is the component of velocity perpendicular to ${\bf{b}}$, and $s$ is a line element parallel to ${\bf{b}}$. Finally, $\gamma$ is the Lorentz factor ($\gamma^2=1/\left(1-v^2/c^2\right)=c^2/\left(c^2-v^2\right)$). Using this factor, we define a relativistic parallel velocity $\upar(=\gamma\vpar)$ for simplicity of notation. 

Further simplifications have been made, by assuming that only electrons or protons are considered in this model; this fixes the rest mass $m_0=m_e=9.1\times10^{-31}\rm{kg}$ and charge $q=e=-1.6022\times10^{-19}\rm{C}$ for electrons, or $m_0=m_p=1.67\times10^{-27}\rm{kg}$ and $q=|e|=1.6022\times10^{-19}\rm{C}$ for protons. In this way, several normalising constants in Eqs.~(\ref{eq:rel_norm}) may be expressed in terms of a normalising electron/proton gyro-frequency, $\omscl(={q\,\bscl}{m_0}^{-1})$. The factor of $\omscl\tscl$ thus plays a key role in controlling the scales at which certain guiding centre drifts become important. 

Relativistic effects not only modify existing terms in the equivalent non-relativistic forms of these equations, but also introduce two new terms in Eq.~(\ref{eq:Rnorm2}) in the direction of $\Eperp$ (i.e. in the ${\bf{b}}\times\ue$ direction). Both of these additional terms are scaled by $\vsclsq/{c^2}$, and as such are purely relativistic.

Finally, several quantities in Eqs.~(\ref{eq:rel_norm}) now also depend on the ratio of perpendicular electric field ($E_\perp$) to the size of the magnetic field ($B$); for a given quantity $H$, $H^{\star}$ and $H^{\star\star}$ are defined as
\[
 H^\star=H\left( 1-\frac{1}{c^2}\frac{{\Eperp}^2}{B^2}\right)^{\frac{1}{2}} , \qquad H^{\star\star}=H\left(1-\frac{1}{c^2}\frac{{\Eperp}^2}{B^2}\right).
\]
These multiplying quantities are dimensionless, i.e. $H^\star$ and $H^{\star\star}$ retain the dimensions of $H$.

We evolve each of Eqns.~(\ref{eq:rel_norm}) in time using a 4th order Runge-Kutta scheme with a variable timestep, subject to the (analytic) electric and magnetic fields outlined in Eqs.~(\ref{eq:B0})-(\ref{eq:B}). A similar approach has been used by, for example \citet{paper:Gordovskyyetal2010b,paper:OskouiNeukirch2014}. We also assume that the separation of spatial/temporal scales between the gyro-motion and global field environment (mentioned in Sec.~\ref{subsec:globalfield}) is sufficient justification for the use of the guiding-centre approximation in this case. This assumption is tested for all orbits (see end of Sect.~\ref{sec:localbehaviour}).

\section{Typical (individual) particle orbits}\label{sec:localbehaviour}
\begin{table}[t]
 	\caption{Individual particle positions}
      \label{tab:party}
      \centering
        \begin{tabular}{l|ccc|cc}\hline\hline
 &\multicolumn{3}{ c|}{Initial pos. (m)}&Remains&Peak	\\
     	&x&y&z& in box?	&energy (eV)			\\
               \hline
 A&$-300$&$-300$&$0$&Y&$2.00$\\
 B&$-100$&$-100$&$0$&N&$52.6$\\
 C&$-20$&$-20$&$0$&N&$6.24\times10^{5}$\\
 D&$-300$&$-300$&$2\times10^{7}$&Y&$2.00$\\
               \hline
           \end{tabular}
\tablefoot{Key to initial positions for individual particles A-D. For reference, also included are the corresponding final state and peak energy for electrons with initial kinetic energies of $2\rm{eV}$ and pitch angle $\theta_{ini}=45\degr$.} 
\end{table}
Before studying the behaviour of many particles in the vicinity of a reconnecting separator, it seems prudent to study several specific examples of particle motion, to identify key aspects of the global behaviour in our calculations. In this section, we will investigate how the behaviour of individual electron and proton orbits varies according to initial position, kinetic energy and pitch angle.
\begin{figure*}
 \centering
 \sbox{\bigleftbox}{%
 \begin{minipage}[b]{.58\textwidth}
  \centering
  \vspace*{\fill}
  \subfloat[Electron paths, initial/final positions, mirror points, interpolated magnetic field lines with the nulls, fan planes and spines included for context.]
  {\label{subfig:Epaths}\resizebox{\textwidth}{!}{\includegraphics{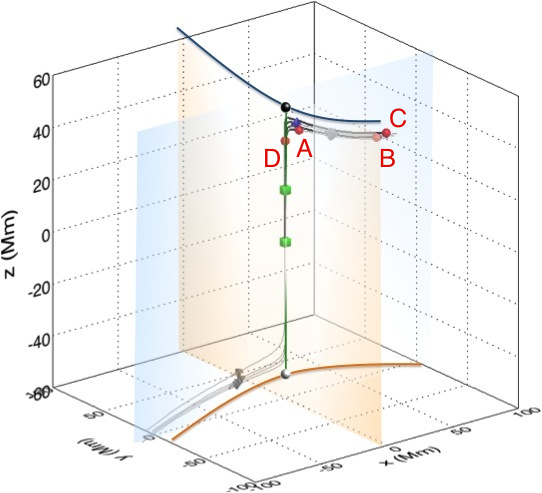}}}
 \end{minipage}%
 }\usebox{\bigleftbox}%
 \begin{minipage}[b][\ht\bigleftbox][s]{.42\textwidth}
  \centering
  \subfloat[Kinetic energy and $\Epar$]
  {\label{subfig:EKEvEpar}\resizebox{\textwidth}{!}{\includegraphics{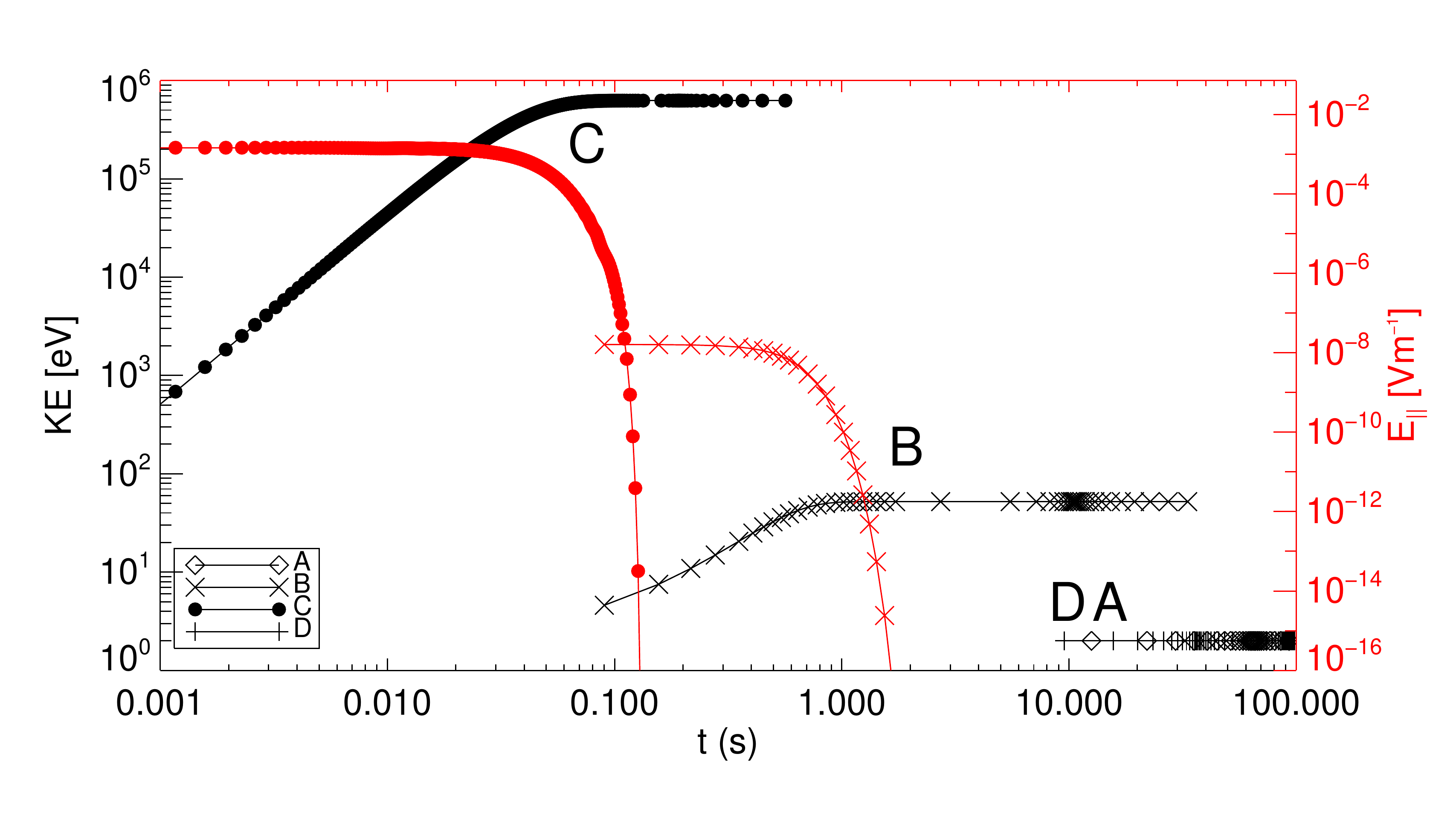}}}
  \vfill
  \subfloat[Normalised $\vpar$ and $|B|$]
  {\label{subfig:EVvB}\resizebox{\textwidth}{!}{\includegraphics{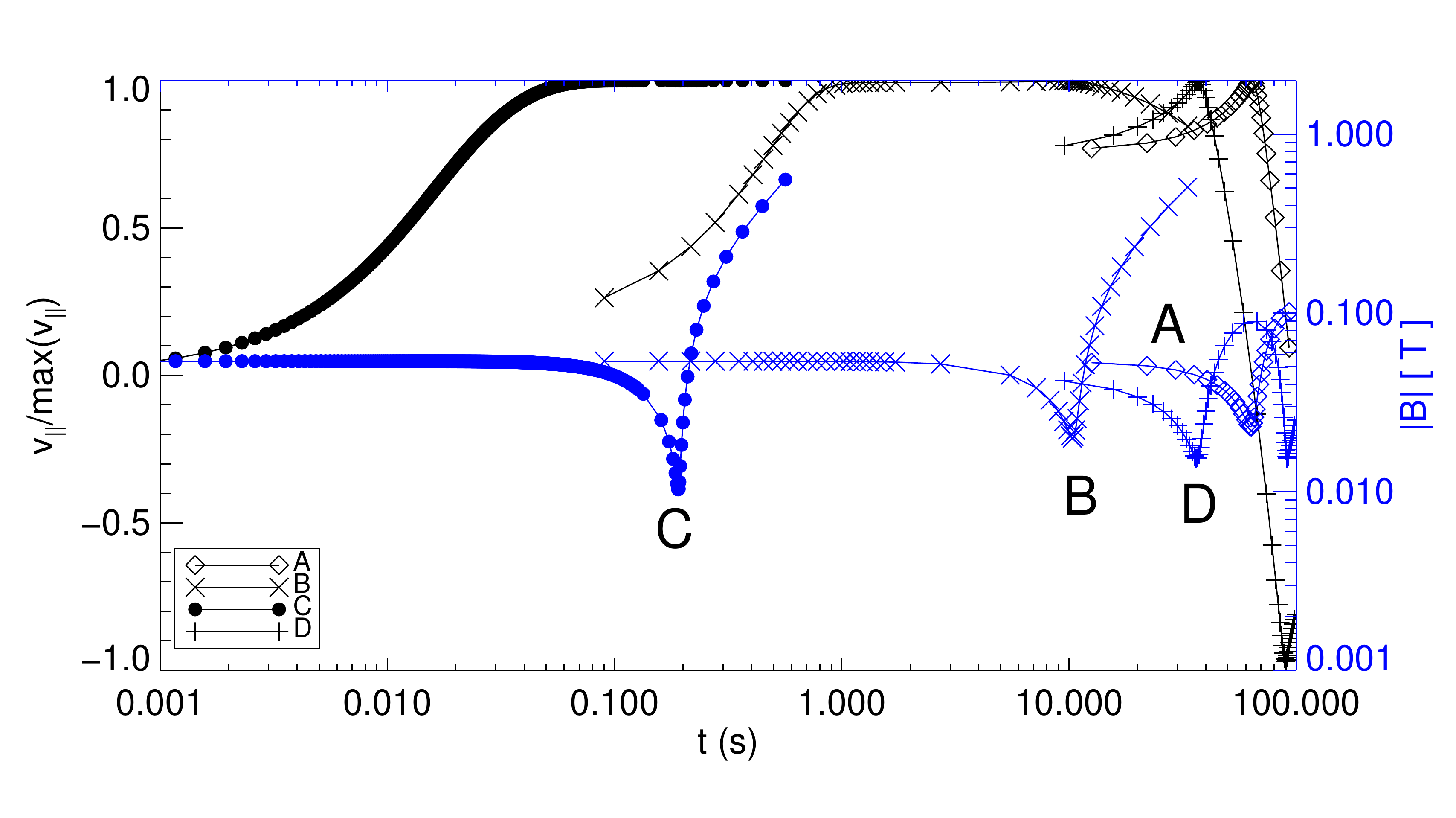}}} 
 \end{minipage}
 \caption{Electron dynamics as a function of time ($KE_{\rm{ini}}=2\rm{eV}$, $\theta_{\rm{ini}}=45\degr$). \protect\subref{subfig:Epaths} The trajectories of four particles in the global separator field, with orbs indicating the starting position (green), final position (red) and any locations where the particles mirror (blue trapezoids); the (thick black) trajectories are overlaid onto local magnetic field lines (grey), with arrows indicating the field orientation. To add context, we also include the nulls and their associated fans and spines, coloured for direct comparison with Fig.~\ref{fig:WS}.  \protect\subref{subfig:EKEvEpar} The change in kinetic energy (black) and parallel electric field (red) experienced by each particle. \protect\subref{subfig:EVvB} A comparison of the (self-normalised) parallel velocity (black) and magnetic field (blue). The initial positions studied here are recorded in Tab.~\ref{tab:party}.}
 \label{fig:indipartsE}
\end{figure*}
\begin{figure*}
 \centering
 \sbox{\bigleftbox}{%
 \begin{minipage}[b]{.565\textwidth}
  \centering
  \vspace*{\fill}
  \subfloat[Proton paths, initial/final positions, mirror points and several interpolated magnetic field lines]
  {\label{subfig:Ppaths}\resizebox{\textwidth}{!}{\includegraphics{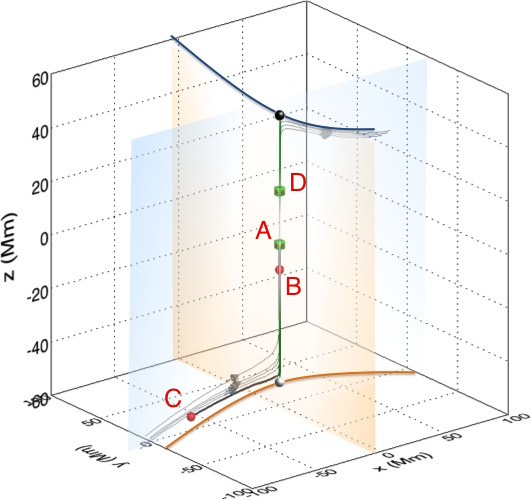}}}
 \end{minipage}%
 }\usebox{\bigleftbox}%
 \begin{minipage}[b][\ht\bigleftbox][s]{.435\textwidth}
  \centering
  \subfloat[Kinetic energy and $\Epar$]
  {\label{subfig:PKEvEpar}\resizebox{\textwidth}{!}{\includegraphics{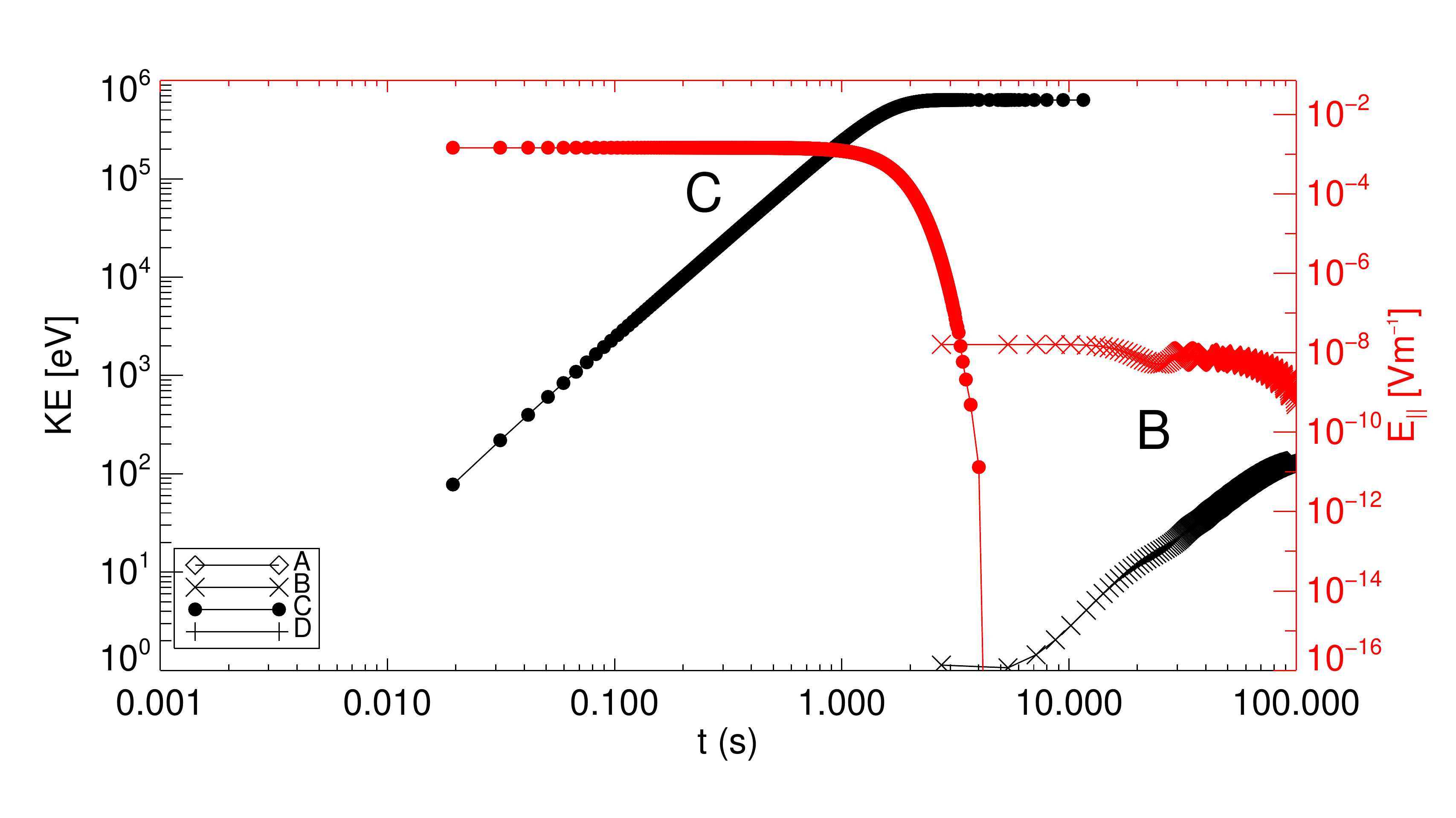}}}
  \vfill
  \subfloat[Normalised $\vpar$ and $|B|$]
  {\label{subfig:PVvB}\resizebox{\textwidth}{!}{\includegraphics{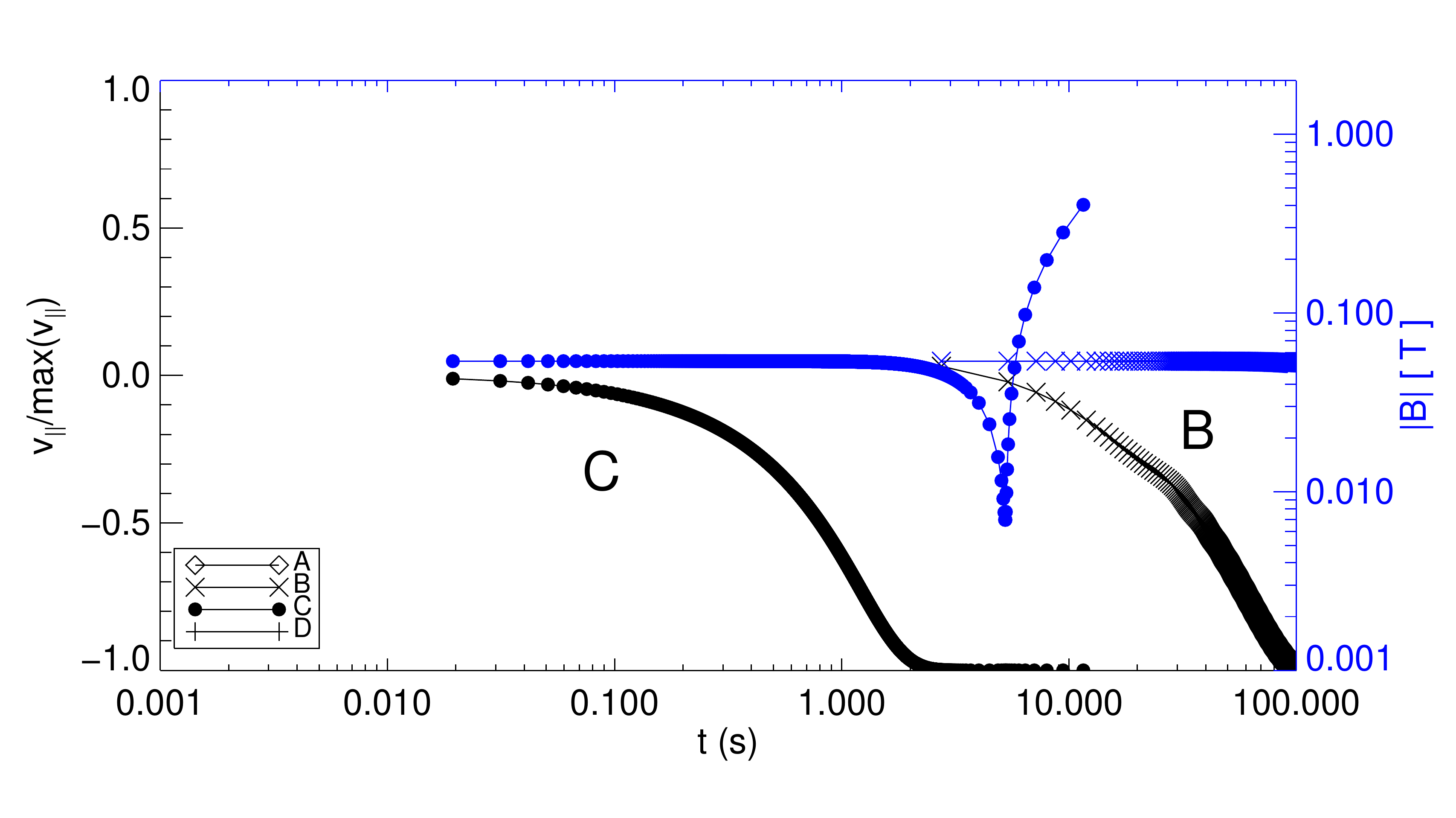}}}
 \end{minipage}
 \caption{Proton dynamics as a function of time ($KE_{\rm{ini}}=2\rm{eV}$, $\theta_{\rm{ini}}=45\degr$). \protect\subref{subfig:Ppaths} The trajectories of four particles in the global separator field, with orbs indicating the starting position (green), final position (red) and any locations where the particles mirror (blue trapezoids); the (thick black) trajectories are overlaid onto local magnetic field lines (grey), with arrows indicating the field orientation. For comparison with Fig.~\ref{fig:WS}, we also include an impression of the fan plane structure and approximate null and spine locations. \protect\subref{subfig:PKEvEpar} The change in kinetic energy (black) and parallel electric field (red) experienced by each particle. \protect\subref{subfig:EVvB} A comparison of the (self-normalised) parallel velocity (black) and magnetic field (blue). The initial positions studied here are recorded in Tab.~\ref{tab:party}.}
 \label{fig:indipartsP}
\end{figure*}
 \begin{figure*}[t]
 \centering
 \subfloat[Electron gyro-radii]{\label{subfig:Egyro}\includegraphics[width=0.49\textwidth]{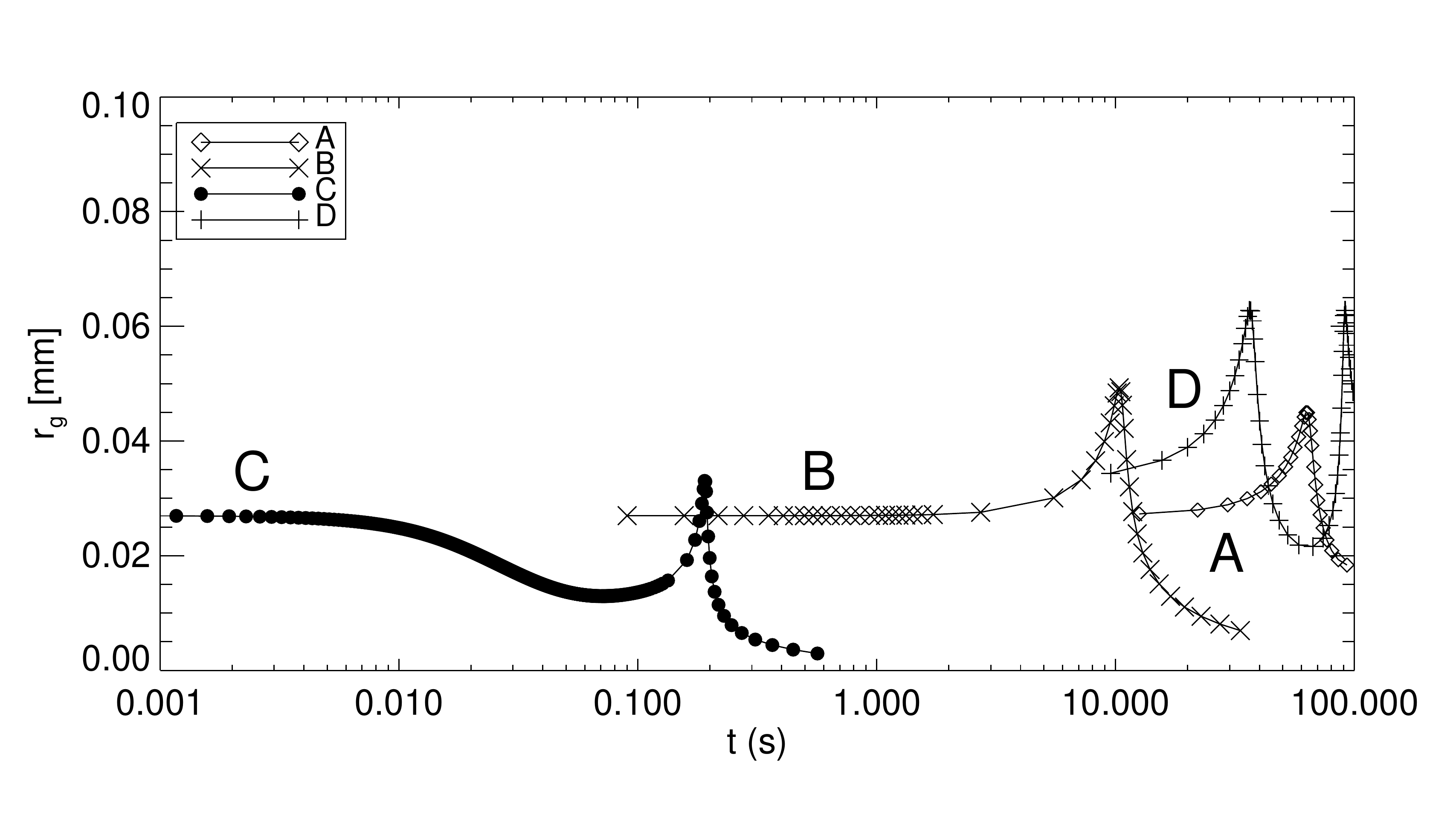}}
 \subfloat[Proton gyro-radii]{\label{subfig:Pgyro}\includegraphics[width=0.49\textwidth]{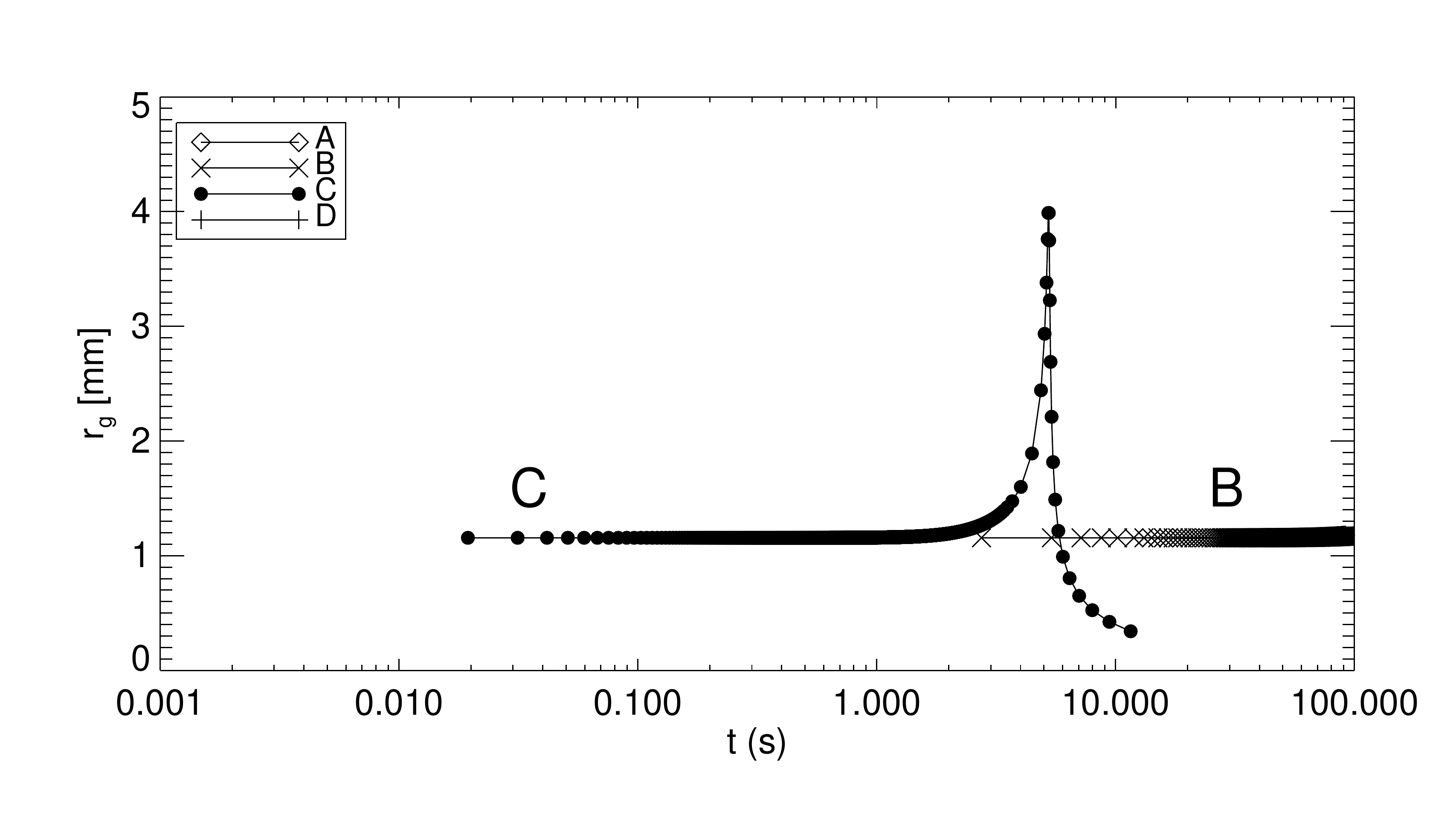}}
 \caption{Evolution of gyro-radius of particles with time, for experiments highlighted by Figs.~\ref{fig:indipartsE}-\ref{fig:indipartsP}. Both electrons \protect\subref{subfig:Egyro} and protons \protect\subref{subfig:Pgyro} have initial positions described in Tab.~\ref{tab:party}, with $KE_{\rm{ini}}=2\rm{eV}$ and $\theta_{\rm{ini}}=45\degr$.}
 \label{fig:gyro}
\end{figure*}

Throughout this investigation, we choose $z_0=5\lscl$ (i.e. a separator of $100\rm{Mm}$ in length) and $l=0.2{z_0}$ (i.e. a flux ring which significantly decays after $10\rm{Mm}$ above/below the vertical midplane). This represents a separator with the current density (and hence reconnection region) localised in a short spatial domain midway along its length. Our first experiment also has the parameter $a=10^{-6}z_0$ (in order to concentrate the flux ring almost exclusively on the separator); this will allow us to examine how the electric field (and the associated magnetic flux ring) affects particles whose initial positions progressively move closer to the separator. At this value, $a=50\rm{m}$; this value is also close to the typical current sheet width value discussed at the end of Sec.~\ref{subsec:globalfield}. 
Therefore, from Eq.~\ref{eq:E}, $|E|\simeq0.1\rm{V\,m^{-1}}$. In comparison to the electric field strengths used in 2D or 2.5D models, our electric field appears to be too small to be relevant for either accelerating particles or reconnecting flux. However, it is important to remember that, here, we are considering a 3D reconnection model with an electric field present over a large distance ($\simeq20\rm{Mm}$). These electric fields lead to reconnection rates which are in line with those found in numerical experiments, as discussed in detail in Sec.~\ref{sec:discussion}. Also, this means particles may experience the accelerating force of an electric field over a considerable distance (time) and, hence, the work done on any given particle by the field is significant. Indeed, we estimate the peak energy gain possible for this electric field configuration to be $8.85\times10^{7}\rm{eV}$ for the present parameters, which is sufficient to accelerate particles up to relativistic speeds, as will be shown below.

\subsection{Effect of varying initial position}\label{subsec:inipos}

We begin by giving each particle a relatively small initial kinetic energy ($2\rm{eV}$). This is a much smaller energy than the typical thermal energy in the solar corona, but we have deliberately reduced this in order to study potentially significant acceleration of particles upon encountering a weak, but extended, parallel electric field. Particles with a small initial kinetic energy are likely to travel much slower, allowing for a better description of the particle behaviour. The initial energy is divided between parallel- and gyro-motion through an initial pitch angle, $\theta_{\rm ini}$; at any given time, the pitch angle $\theta$ is given as
\[
 \theta=\arccos{\left(\frac{\vpar}{v_{\rm{tot}}} \right) },
\]
where $\vpar$ is the component of the guiding centre velocity parallel to the local magnetic field, and $v_{\rm{tot}}$ is the total particle velocity. For this investigation (and indeed for the majority of the global simulations in Sec.~\ref{sec:globalbehaviour}), we will use $\theta_{\rm{ini}}=45\degr$.

Each fan plane acts as a boundary layer between two topologically distinct domains. We seek to uncover the general characteristics of particle behaviour in a given topological domain. By initialising four particles at equal distances from these topological boundaries (e.g. by choosing initial positions where $x=y$) we hope to avoid non-generic effects which might arise by placing particles close to or on these boundaries. Three of these initial positions are situated in the vertical midplane, at a range of distances from the separator. For reference, we label these particles A-C (full details are given in Tab.~\ref{tab:party}). A fourth particle is similarly positioned equidistant from both fan planes, but placed $20\rm{Mm}$ above the vertical midplane and outside the reconnection region (for comparison with A-C); we label this particle D. Both the particle trajectories, and a comparison of properties over the time of calculation for each particle, are displayed in Fig.~\ref{fig:indipartsE} for electrons and Fig.~\ref{fig:indipartsP} for protons. 
\begin{figure*}[t]
 \centering
 \begin{tabular}{ccc}
  \subfloat[Change in final position with $\theta$ (location A, $2\rm{eV}$)] {\label{subfig:Apangle}\resizebox{0.4\textwidth}{!}{\includegraphics{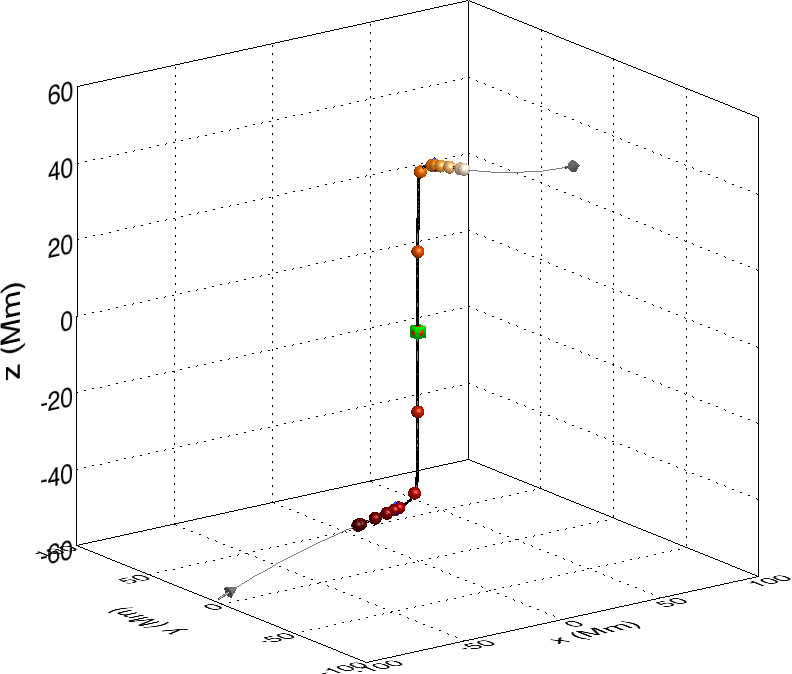}}}	&
  \resizebox{0.12\textwidth}{!}{\includegraphics{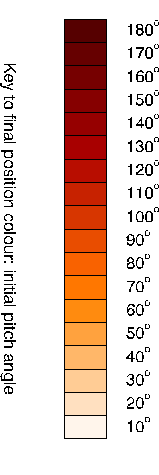}} 				&
  \subfloat[Change in final position with $\theta$ (location A, $200\rm{eV}$)] {\label{subfig:Apangle200eV}\resizebox{0.4\textwidth}{!}{\includegraphics{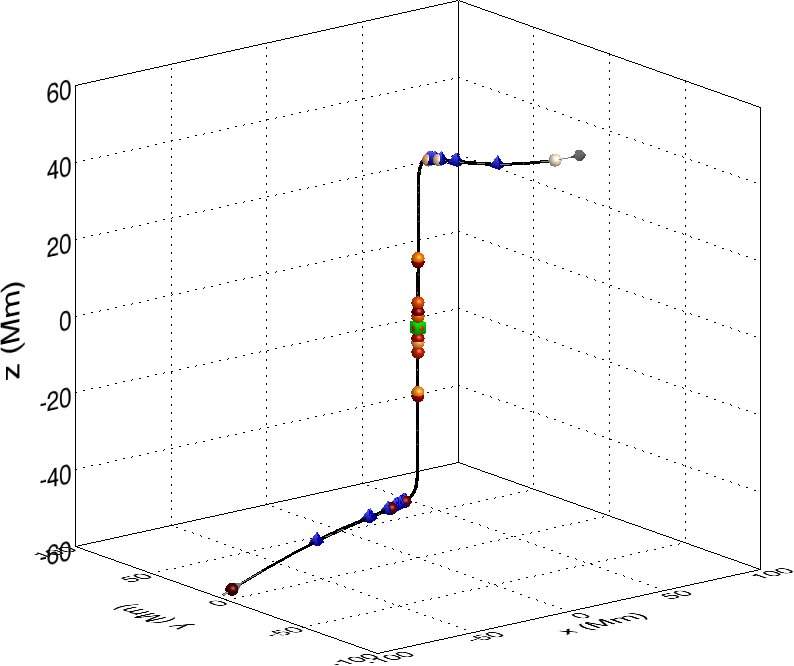}}} \\
  \subfloat[Change in final position with $\theta$ (location C, $2\rm{eV}$)] {\label{subfig:Cpangle}\resizebox{0.4\textwidth}{!}{\includegraphics{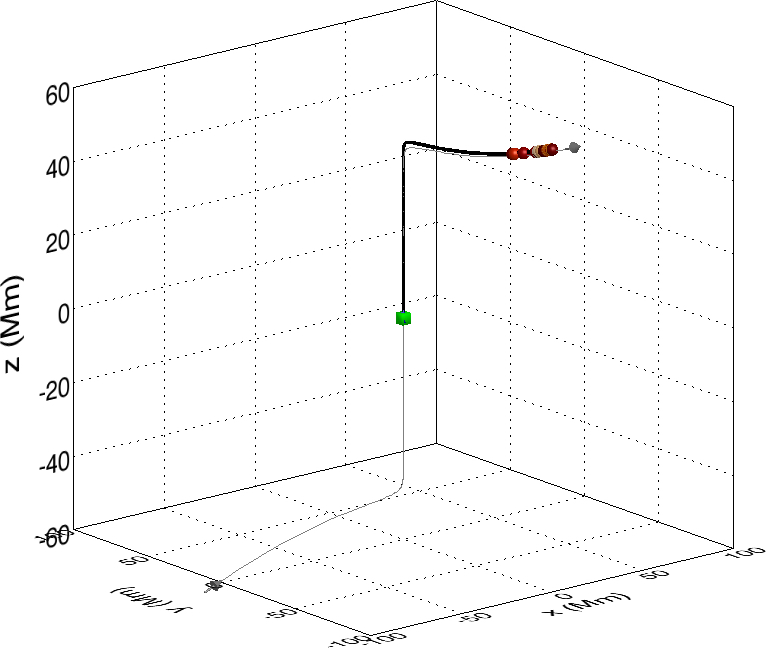}}}	&
  \resizebox{0.12\textwidth}{!}{\includegraphics{24366fg5a1}} 				&
  \subfloat[Change in final position with $\theta$ (location C, $200\rm{eV}$)] {\label{subfig:Cpangle200eV}\resizebox{0.4\textwidth}{!}{\includegraphics{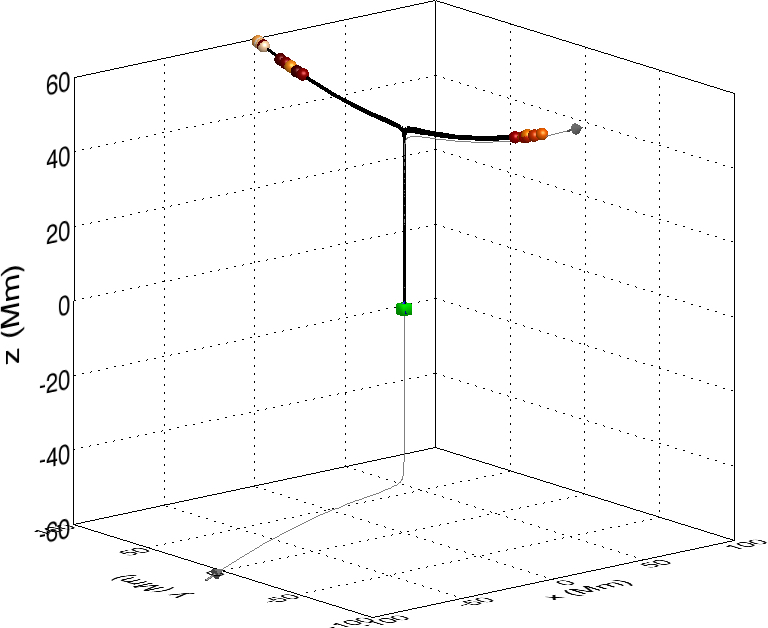}}} \\
 \end{tabular}
 \caption{Investigation into effect of pitch angle and initial energy. \protect\subref{subfig:Apangle} Variation of the particle trajectories at position A (specifically their final positions) with initial pitch angle at an energy of $2\rm{eV}$. \protect\subref{subfig:Apangle200eV} The same result for initial energies of $200\rm{eV}$. \protect\subref{subfig:Cpangle} and \protect\subref{subfig:Cpangle200eV} Trajectories and end-point locations for particles beginning at position C (see Tab.~\ref{tab:party}). In all cases the final positions are colour-coded, depending on initial pitch angle (for key, see colour bar).}
 \label{fig:varythetaandke}
\end{figure*}

We begin by examining the behaviour of electrons, using direct references to aspects of Fig.~\ref{fig:indipartsE}. Of the electrons which begin in the vertical midplane, 
A is the furthest from the separator, while C is the closest; this change in distance from the separator is the controlling factor for electron behaviour in this experiment.

While electron A experiences no electric field over the time of the calculation (note, in Fig.~\ref{subfig:EKEvEpar} no red curve is visible for A and D), electrons at positions B and C experience moderate and strong{\footnote{`Strong' in this context (and throughout the paper) means an electric field that is large in comparison to the peak electric field in the model, which, as already discussed, is considered small compared to the electric fields found, for instance, in 2D steady-state reconnection models.}} electric fields, respectively. Fig.~\ref{subfig:Epaths} shows that this causes B and C to be accelerated upwards along the separator and out on a trajectory similar to that of the spine of the upper null, terminating at the edge of the numerical box. A and D also travel upwards; while they follow similar trajectories, these electrons 
travel at much slower speeds and end the simulation much closer to the upper null. 
\begin{figure*}[t]
 \centering
 \subfloat[Field config, $t=0$]{\label{subfig:twistt0}\includegraphics[height=0.33\textheight]{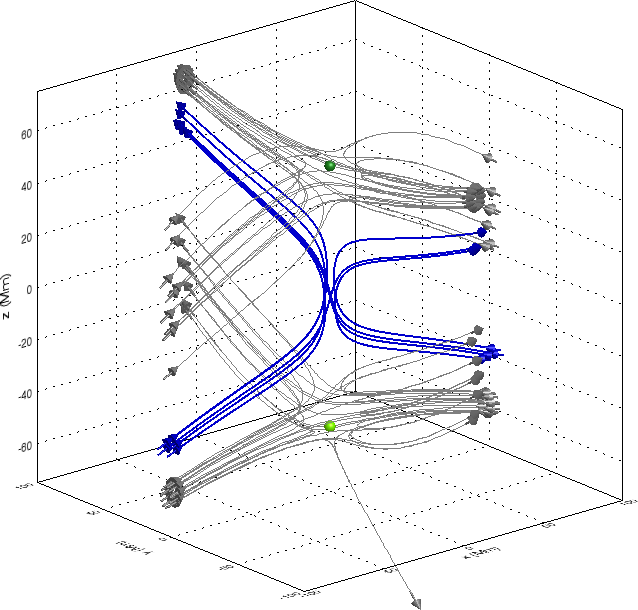}}
 \subfloat[Field config, $t=100\rm{s}$]{\label{subfig:twistt100}\includegraphics[height=0.33\textheight]{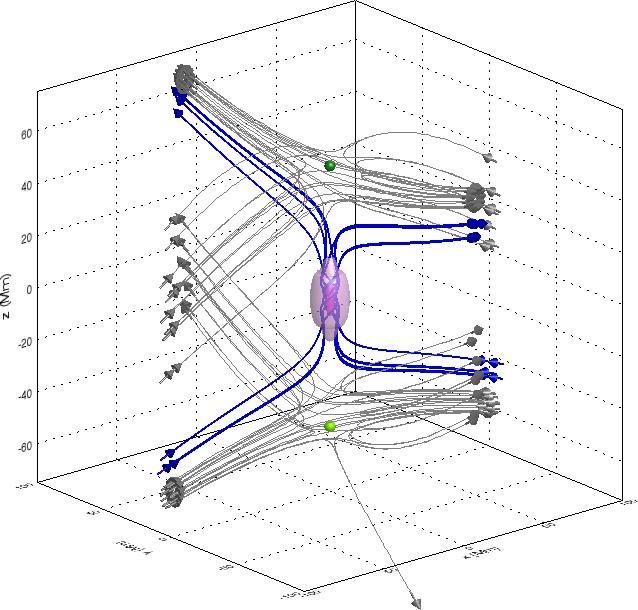}}
 \caption{Illustration of twisting magnetic field, for specific field lines given by Eqs.~\ref{eq:B0}-\ref{eq:B}, at \protect\subref{subfig:twistt0} $t=0$ and \protect\subref{subfig:twistt100} $t=100\rm{s}$. 
 While the grey field lines illustrate the field lines originating near the spines of either null (green orbs), the blue field lines highlight field lines close to the separator and the pink isosurfaces highlight regions of parallel current, $|j_{||}|$ above approximately $10\%$ and $50\%$ of the peak value. In these examples $a=0.1z_0$, to enhance visibility of the twisting of (blue) field lines around the separator.}
 \label{fig:twist}
\end{figure*}

The extended region of intense electric field experienced by C causes rapid acceleration to a relativistic parallel velocity in a very short time; C exits the numerical box after only $0.8\rm{s}$. At this point in the simulation, Fig.~\ref{subfig:EKEvEpar} shows that the kinetic energy of C has grown to $0.624\rm{MeV}$ ($0.07\%$ of maximum possible energy gain, $8.85\times10^7$eV), while its parallel velocity is approximately $269\rm{Mm\,s}^{-1}$ ($0.89c$). Electron B also leaves the box after around $35\rm{s}$; as the electric field it experiences is much weaker, the final kinetic energy of B is $52.6\rm{eV}$, with a corresponding parallel velocity of $4.3\rm{Mm\,s}^{-1}$ ($0.013c$). Electrons A and D remain in the numerical box for all time, and only achieve peak parallel velocities of $780\rm{km\,s}^{-1}$ and $800\rm{km\,s}^{-1}$ respectively, while both retain a kinetic energy of $2\rm{eV}$.

Electrons A-C all travel almost exactly parallel to the magnetic field throughout the calculation; while guiding-centre drifts are accounted for in the model, in the present setup they have 
magnitudes of $\rm{ms}^{-1}$ speeds, which (for the scales plotted in Fig.~\ref{subfig:Epaths}) are negligible compared to the speed of parallel motion. The sharpest change in parallel velocity (not directly caused by the electric field) occurs for electron D, as it encounters a mirror point close to the upper null (shown as a blue pyramid in Fig.~\ref{subfig:Epaths}). Approximately $60\rm{s}$ into the simulation, D experiences a significant increase in magnetic field strength, which causes the particle velocity to reverse sign (crosses in Fig.~\ref{subfig:EVvB}). This is the reason that the final position of D is closer to the vertical midplane than the other electrons (see Fig.~\ref{subfig:Epaths}).

We now turn our attention to the proton behaviour, exhibited in Fig.~\ref{fig:indipartsP}. Due to the difference in charge ($q=|e|$) and mass ($m_p\simeq1836me_e$), we would broadly expect protons to travel in the opposite direction to electrons at lower speeds; this behaviour is readily apparent from Fig.~\ref{subfig:Ppaths}, where the particle trajectories are shorter and in the opposite direction to those seen in Fig.~\ref{subfig:Epaths}. We also observe that the normalised parallel velocity in Figs.~\ref{subfig:EVvB} and~\ref{subfig:PVvB} almost always differs in sign, indicating that protons and electrons travel in opposite directions in this experiment (unless mirror points are encountered). Once again, particles placed at positions A and D do not encounter any (significant) electric field. With an initial kinetic energy of $2\rm{eV}$ at an initial pitch angle of $45\degr$, each proton has an approximate initial parallel velocity of $13\rm{km\,s}^{-1}$, while an equivalent electron would begin the experiment with $\vpar\simeq593\rm{km\,s}^{-1}$. This difference in speed is due to the mass difference of the particles. Thus, protons which do not encounter the electric field are not seen to travel over the course of the simulation (at the normalising values chosen for this experiment). This is the reason Fig.~\ref{subfig:Ppaths} shows that protons A and D have not moved from their initial positions (green cube and red orbs overlap).

Protons B and C encounter moderate and strong electric fields. Due to their larger mass, they are more slowly accelerated than electrons; this is clear from the behaviour of proton B in Fig.~\ref{subfig:Ppaths}, which has travelled only a fraction of the distance covered by an electron starting from the same position in Fig.~\ref{subfig:Epaths}. Proton B achieves a peak kinetic energy of $13.6\rm{eV}$, corresponding to a peak parallel velocity of $161\rm{km\,s}^{-1}$; after $100\rm{s}$, this proton has yet to leave the reconnection region and is still accelerating/gaining energy. Also noteworthy is the electric field experienced by proton B, according to Fig.~\ref{subfig:PKEvEpar}; towards the end of the experiment, the electric field strength experienced by B (red crosses) becomes oscillatory in nature. This will be studied further in Sec.~\ref{subsec:inikeandtheta}. Proton C, which feels the strongest direct acceleration, achieves a peak energy of $0.634\rm{MeV}$, and exits the computational domain with a peak parallel velocity of $11\rm{Mm\,s}^{-1}$ ($0.037c$) after $10.5\rm{s}$. Proton C also takes longer to leave the computational domain than electron C (comparing Figs.~\ref{subfig:EKEvEpar} and~\ref{subfig:PKEvEpar}) and does so via field lines close to the spine of the lower null; as expected, this is due to the difference in proton/electron mass and charge.

Finally, Fig.~\ref{fig:gyro} provides an estimate of the gyro-radius for both electrons and protons over the course of the simulation. For relativistic cases, the gyro-radius of the particle is determined via
\begin{equation}
 r_g=\frac{p_\perp}{|q|B}=\frac{1}{|q|}\sqrt{\frac{2m_0\mu_r}{B}}.
 \label{eq:gyroradius}
\end{equation}
Due to the presence of the rest mass ($m_0$) in this expression, we expect that protons will have a larger gyroradius than electrons (by a factor of $\sqrt{m_p/m_e}\simeq42.85$). 
Expression (\ref{eq:gyroradius}) also shows that particles which come very close to either null-point will exhibit a gyro-radius which increases rapidly. Figure~\ref{fig:gyro} indeed shows that the largest gyro-radius (of approximately $4\rm{mm}$) is achieved by proton C, due to its encounter with the region around the lower null where magnetic field strength decreases. The evolution of electron gyro-radii (Fig.~\ref{subfig:Egyro}) is complicated by the presence of magnetic mirror points, which ultimately cause the particles to re-encounter the region of lower magnetic field strength near a null. As a result, multiple peaks may be observed in the gyro-radii of some particles, e.g. electron D in Fig.~\ref{subfig:Egyro}. 

For both protons and electrons, even in the cases of particle acceleration to relativistic parallel velocities, the gyro-radii displayed in Fig.~\ref{fig:gyro} remain several orders of magnitude smaller than the (metre) length-scales of the simulation. As the scales of both guiding centre and global field models remain well separated for all time, our use of the guiding-centre approximation to study particle behaviour in this environment is well justified.

\begin{figure*}[t]
 \centering
 \subfloat[$2\rm{eV}$; electron paths] {\label{subfig:a1e-6tracks}\includegraphics[height=0.25\textheight]{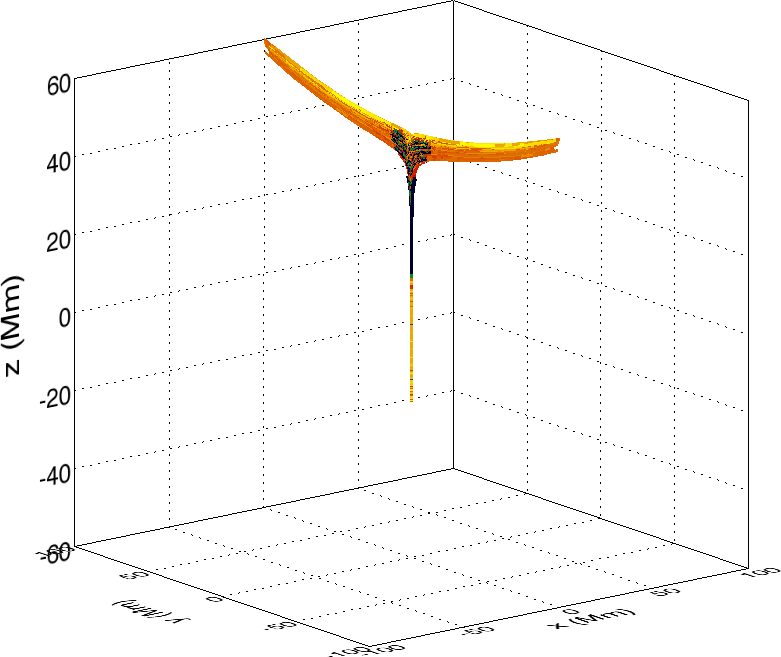}\includegraphics[height=0.25\textheight]{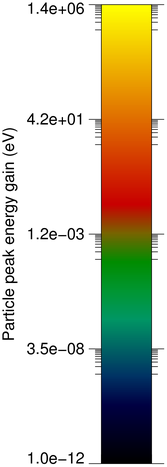}}
 \subfloat[$2\rm{eV}$; ini. positions ($-0.4\rm{km}<x,y<0.4\rm{km}$)]{\label{subfig:a1e-6inipos}\includegraphics[height=0.25\textheight]{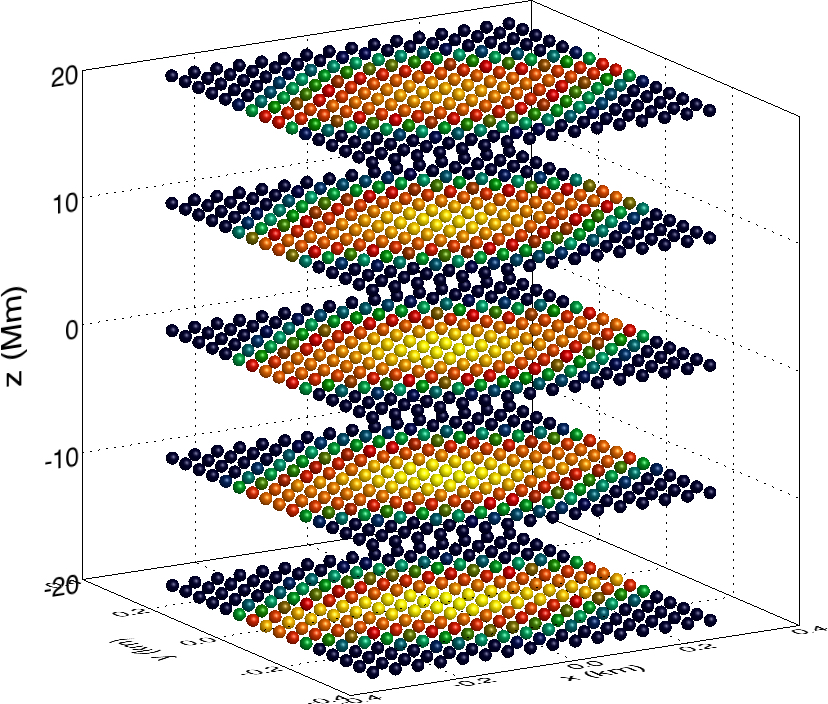}}\\
 \subfloat[$200\rm{eV}$; electron paths] {\label{subfig:a1e-6tracks2}\includegraphics[height=0.25\textheight]{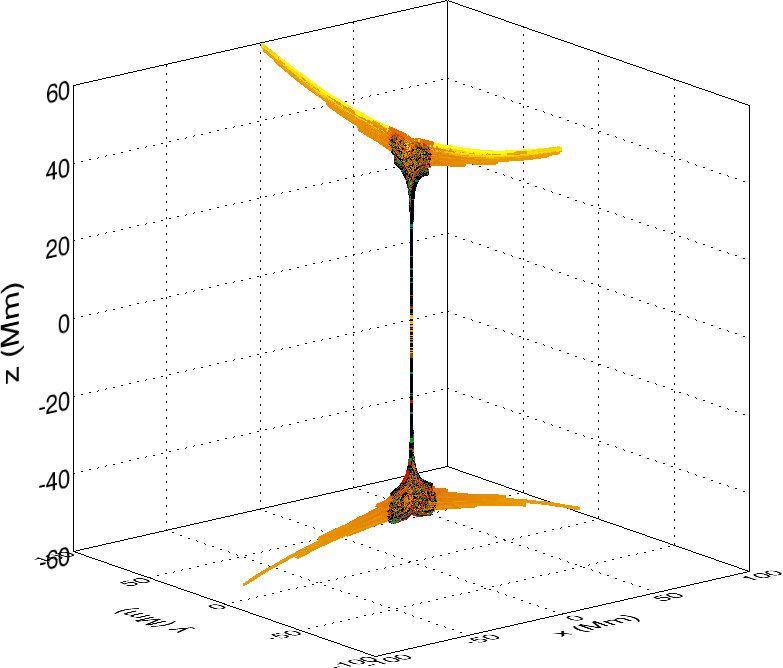}\includegraphics[height=0.25\textheight]{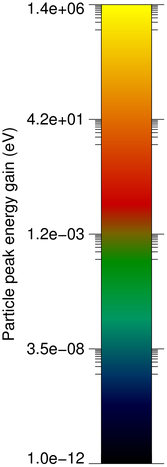}}
 \subfloat[$200\rm{eV}$; ini. positions ($-0.4\rm{km}<x,y<0.4\rm{km}$)]{\label{subfig:a1e-6inipos2}\includegraphics[height=0.25\textheight]{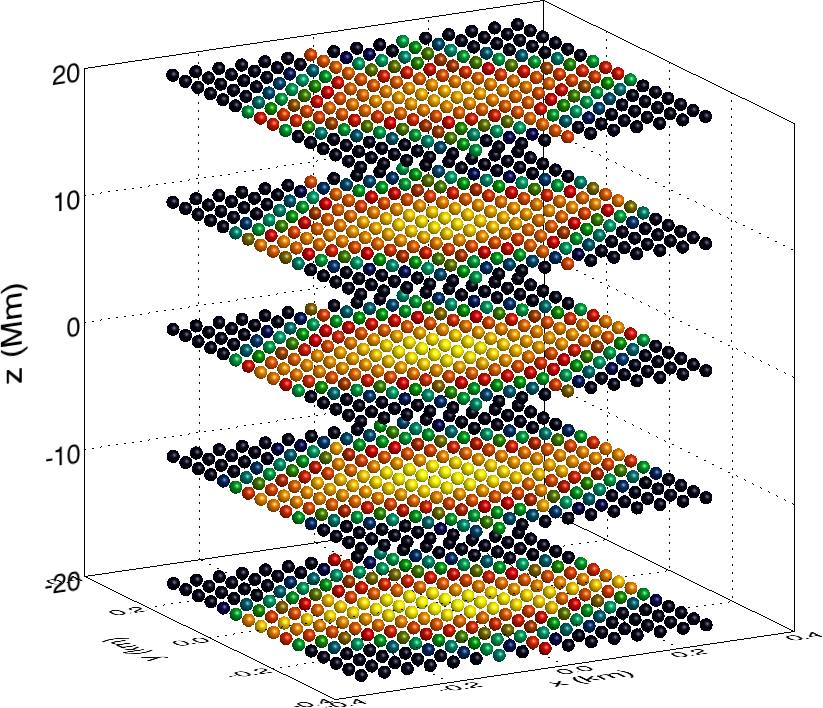}}
 \caption{$a=10^{-6}z_0$; Electron trajectories \protect\subref{subfig:a1e-6tracks} and initial positions \protect\subref{subfig:a1e-6inipos} for $2\rm{eV}$ particles with initial pitch angle $45\degr$. \protect\subref{subfig:a1e-6tracks2} and \protect\subref{subfig:a1e-6inipos2} are the same, but for $200\rm{eV}$ particles. The colour of each particle or track identifies the peak kinetic energy gain of the particle during the simulation (see colour bar).}
 \label{fig:a1e-6}
\end{figure*}

\begin{figure*}[t]
 \centering
 \subfloat[$2\rm{eV}$; Proton paths] {\label{subfig:a1e-6ptracks}\includegraphics[height=0.25\textheight]{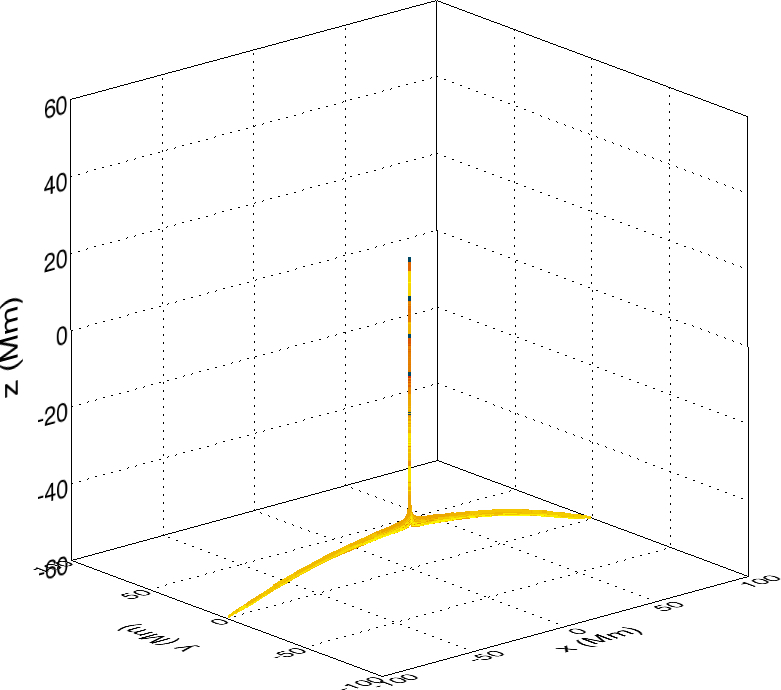}\includegraphics[height=0.25\textheight]{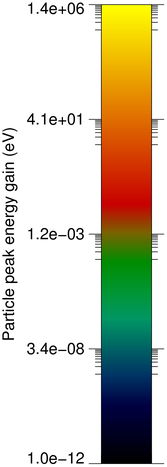}}
 \subfloat[$2\rm{eV}$; ini. positions ($-0.4\rm{km}<x,y<0.4\rm{km}$)]{\label{subfig:a1e-6pinipos}\includegraphics[height=0.25\textheight]{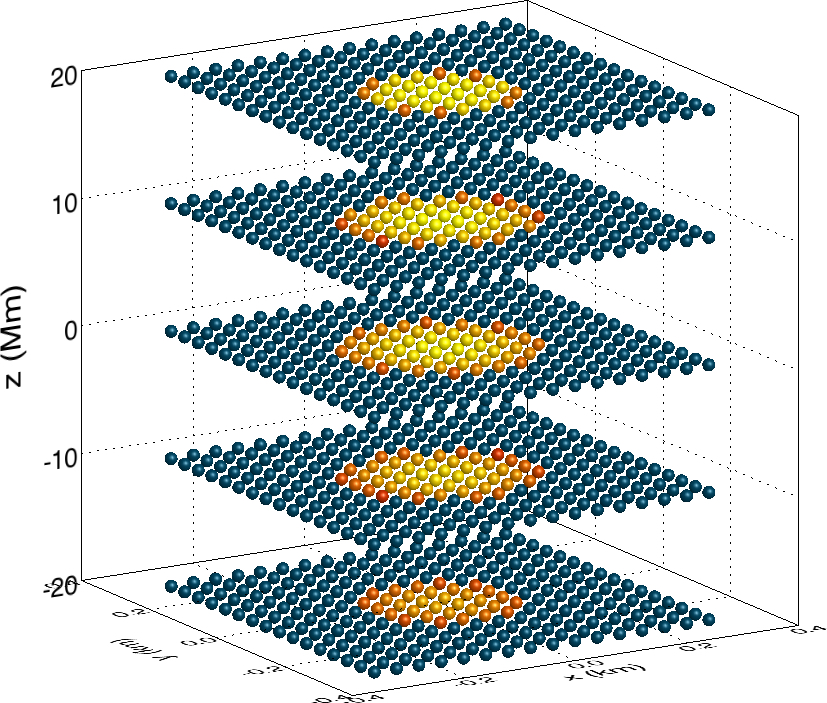}}\\
 \subfloat[$200\rm{eV}$; Proton paths] {\label{subfig:a1e-6ptracks2}\includegraphics[height=0.25\textheight]{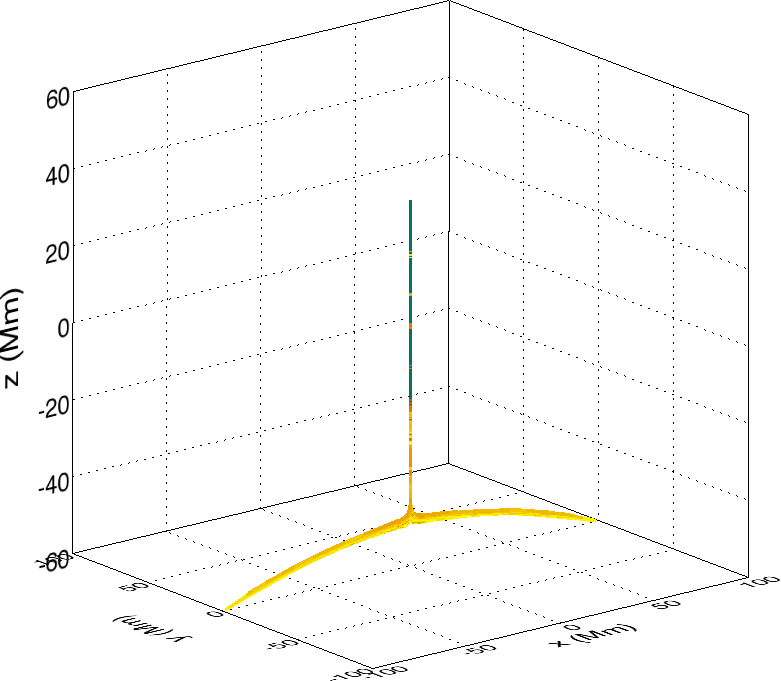}\includegraphics[height=0.25\textheight]{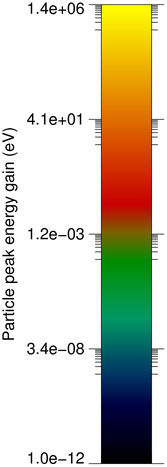}}
 \subfloat[$200\rm{eV}$; ini. positions ($-0.4\rm{km}<x,y<0.4\rm{km}$)]{\label{subfig:a1e-6pinipos2}\includegraphics[height=0.25\textheight]{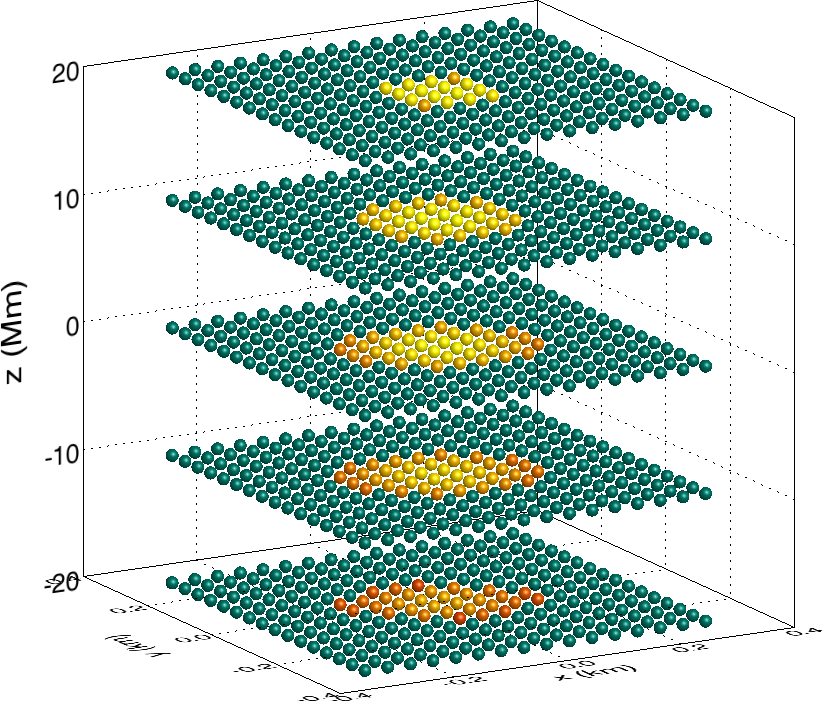}}
 \caption{$a=10^{-6}z_0$; Proton trajectories \protect\subref{subfig:a1e-6ptracks} and initial positions \protect\subref{subfig:a1e-6pinipos} for $2\rm{eV}$ particles with initial pitch angle $45\degr$. \protect\subref{subfig:a1e-6ptracks2} and \protect\subref{subfig:a1e-6pinipos2} are the same, but for $200\rm{eV}$ particles. The colour of each particle or track identifies the peak kinetic energy gain of the particle during the simulation (see colour bar).}
 \label{fig:a1e-6p}
\end{figure*}

\subsection{Effect of varying initial kinetic energy and pitch angle}\label{subsec:inikeandtheta}
Until now, we have focused on the effect of the initial position of the particles on their behaviour. We now turn our attention to the remaining two initial conditions for each particle; kinetic energy and pitch angle. We still hold our initial particle positions fixed at the values used in the previous experiment (given in Tab.~\ref{tab:party}), and continue to refer to these particles by their initial position (i.e. particle A is initialised at location A, etc).

We consider the behaviour of individual particles starting from two initial positions, A and C (in order to compare behaviour with/without the influence of the electric field), for a range of pitch angles from $0-180\degr$, and for two different initial kinetic energy values, $2\rm{eV}$ and $200\rm{eV}$. For brevity, we summarise the results of this investigation for electrons in Fig.~\ref{fig:varythetaandke}, and discuss how the results differ when replacing electrons with protons.

By varying the initial pitch angle at each starting position, we affect how the initial kinetic energy is distributed between the parallel and gyroscopic motion. A small/extremely large initial pitch angle ($0/180\degr$) will cause the majority of the initial $2\rm{eV}$ energy to go towards moving each particle in a parallel/anti-parallel sense along the magnetic field; initial pitch angles close to $90\degr$ will see the majority of the particle energy go towards the gyro-velocity.
\begin{figure}[t]
 \centering
 \includegraphics[height=0.25\textheight]{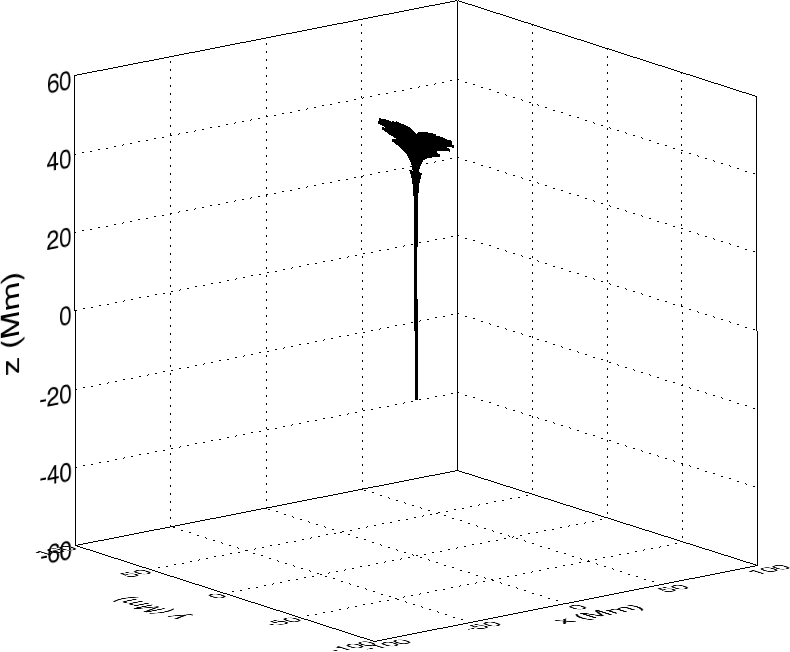}
 \caption{Electron paths for $a=10^{-7}z_0$, with same initial positions as Fig.~\ref{fig:a1e-6}. The peak kinetic energy of all particles remains equal to the initial energy, hence their identical colour (black).}
 \label{fig:a1e-7samepos}
\end{figure}

Figure~\ref{subfig:Apangle} illustrates how the final position of electrons initially located at A vary with pitch angle. This figure shows that (as expected) small pitch angles divert the majority of the initial kinetic energy into motion parallel to the magnetic field. As the pitch angle of electron A grows from zero, the amount of parallel energy available to transport the particle along the field is gradually reduced. Thus the initial electron velocity progressively reduces with pitch angle (with a minimum at $90\degr$), and therefore electron A travels shorter distances from the initial position in the time available. A switch-over occurs as the pitch angle passes through $90\degr$ in Fig.~\ref{subfig:Apangle}; while at exactly $90\degr$, all the kinetic energy goes towards the particle gyro-velocity, and hence the particle position remains at the initial position. This is seen in Fig.~\ref{subfig:Apangle} as a red orb (final position) in the same location as the green cube (initial position). As the pitch angle grows beyond $90\degr$, the parallel velocity of the particle begins to act in the opposite direction (now anti-parallel to the magnetic field). Pitch angles which approach $180\degr$ allow particles to again travel larger distances from the initial position, but in the opposite direction from particles with pitch angles close to $0\degr$. Once more this is seen in Fig.~\ref{subfig:Apangle} in the particle trajectories which travel down along the separator, in the opposite direction to the magnetic field. The same effect is also recovered for protons, but in the opposite direction (due to the difference in charge), and with a greatly reduced velocity/distance travelled (due to the increase in particle mass).

\begin{figure*}[t]
 \centering
 \subfloat[Electron trajectories]{\label{subfig:a1e-7tracks}\includegraphics[height=0.25\textheight]{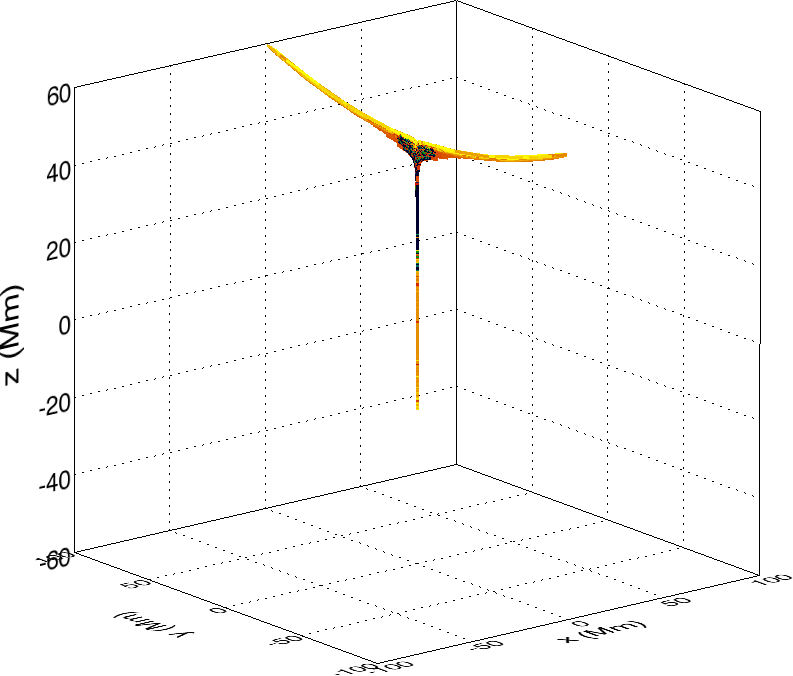}\includegraphics[height=0.25\textheight]{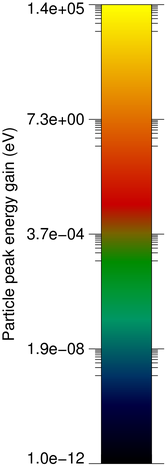}}
 \subfloat[Initial positions ($-40\rm{m}<x,y<40\rm{m}$)]{\label{subfig:a1e-7inipos}\includegraphics[height=0.25\textheight]{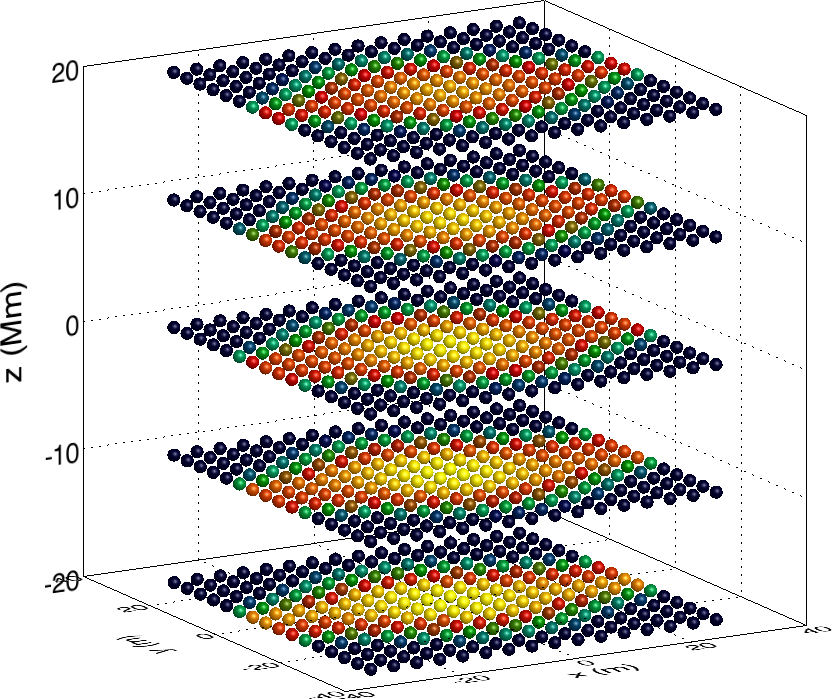}}
 \caption{$a=10^{-7}z_0$; Electron paths \protect\subref{subfig:a1e-7tracks} and initial positions \protect\subref{subfig:a1e-7inipos} for particles with $2\rm{eV}$ initial energy and pitch angle $45\degr$. The colour of each particle or track identifies the peak kinetic energy of the particle during the simulation (see colour bar).}
 \label{fig:a1e-7}
\end{figure*}

\begin{figure*}[t]
 \centering
 \subfloat[Proton trajectories] {\label{subfig:a1e-7ptracks}\includegraphics[height=0.25\textheight]{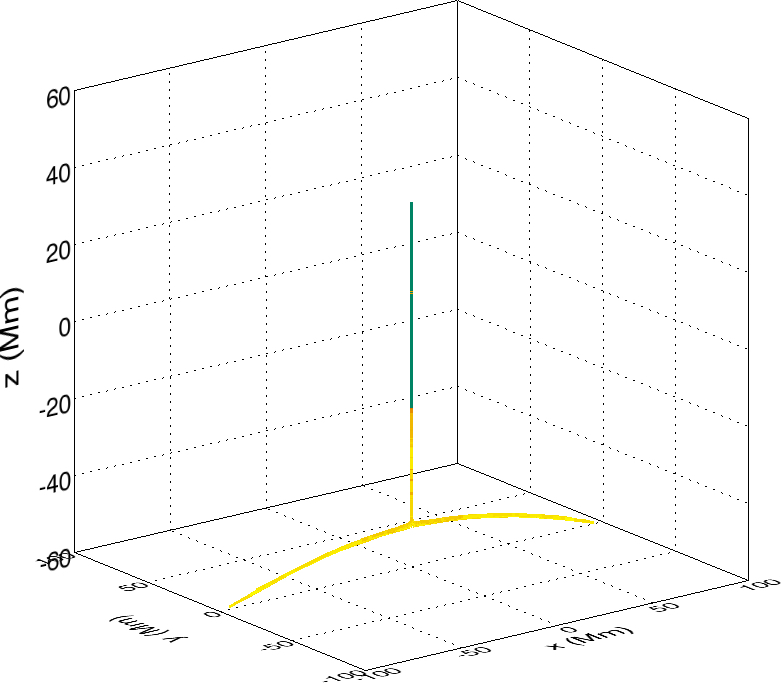}\includegraphics[height=0.25\textheight]{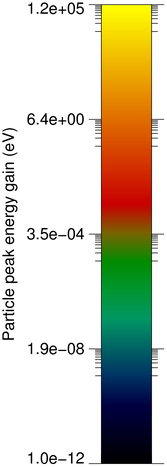}}
 \subfloat[Initial positions ($-40\rm{m}<x,y<40\rm{m}$)]{\label{subfig:a1e-7pinipos}\includegraphics[height=0.25\textheight]{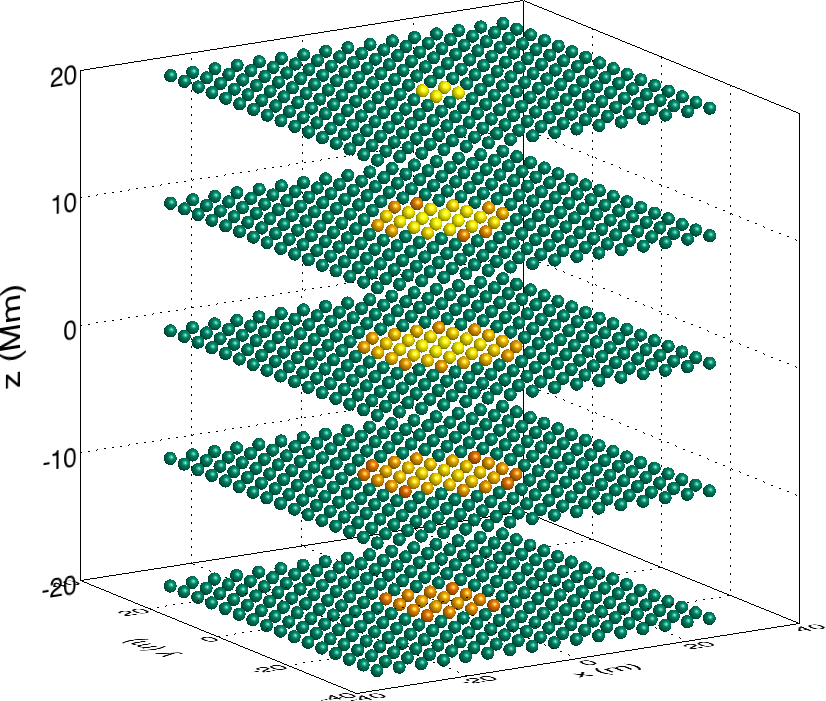}}
 \caption{$a=10^{-7}z_0$; Proton paths \protect\subref{subfig:a1e-7ptracks} and initial positions \protect\subref{subfig:a1e-7pinipos} for particles with $2\rm{eV}$ initial energy and pitch angle $45\degr$. The colour of each particle or track identifies the peak kinetic energy of the particle during the simulation (see colour bar).}
 \label{fig:a1e-7p}
\end{figure*}

We also perform an identical experiment, but with an initial energy of $200\rm{eV}$, in order to assess the impact of initial kinetic energy on particle behaviour; the results of this study are illustrated in Fig.~\ref{subfig:Apangle200eV} (for electrons). At the larger value of initial kinetic energy, the simple pattern of particle behaviour with initial pitch angle (seen in Fig.~\ref{subfig:Apangle}) is no longer present. Instead, Fig.~\ref{subfig:Apangle200eV} shows that increasing the initial kinetic energy causes each electron to more readily encounter magnetic mirror points. The location of the mirror points encountered depends on how the initial $200\rm{eV}$ energy is distributed between parallel and gyro-motion; pitch angles close to $0\degr/180\degr$ not only cause electrons to have a large initial parallel velocity component, but also effectively reduce the particle gyro-radius, meaning that much larger magnetic field strengths are required to mirror these electrons. Conversely, particles with pitch angles close to $90\degr$ not only travel slowly along the magnetic field, but also maintain a larger gyro-radius requiring much weaker magnetic field strengths to cause them to mirror. The same is true for protons, again noting that proton gyro-radii are larger by a factor of $\sqrt{m_e/m_p}$; protons will readily mirror upon encountering regions of increasing magnetic field strength, however it takes longer for them to reach these regions compared to electrons.

Our final stage of this survey concerns how the particle behaviour demonstrated by Figs.~\ref{subfig:Apangle} and~\ref{subfig:Apangle200eV} is altered by the presence of the electric field. We therefore repeated the same experiments for particles initially placed at position C; these results can be seen in Figs.~\ref{subfig:Cpangle} and~\ref{subfig:Cpangle200eV}.

Beginning with initial $2\rm{eV}$ energies, Fig.~\ref{subfig:Cpangle} illustrates that the electric field now dominates the particle motion. All particles are rapidly accelerated along field lines close to a single spine of the upper null, irrespective of pitch angle, and achieve peak energies of $0.62\rm{MeV}$ in less than a second. If the initial energy is increased to $200\rm{eV}$, Fig.~\ref{subfig:Cpangle200eV} shows a curious change in behaviour; while particles with initial pitch angles close to $90\degr$ continue to behave as in the $2\rm{eV}$ case, we find that particles with small ($0-63\degr$) or large ($127-180\degr$) initial pitch angles leave the numerical box close to the \emph{opposite} spine of the upper null than they left from before.

In order to establish the reason for this result, a more detailed analysis of the magnetic field evolution was undertaken. In Fig.~\ref{fig:twist}, we illustrate the magnetic field configuration given by Eqs.~\ref{eq:B0}-\ref{eq:B} at the beginning of the experiment ($t=0$ in Fig.~\ref{subfig:twistt0}) and at the end ($t=100\rm{s}$, Fig.~\ref{subfig:twistt100}), for $a=0.1z_0$ (in order to emphasise any differences on a large scale). Each magnetic field line in the image was calculated using a simple ODE solving routine for our chosen fields. From Fig.~\ref{fig:twist}, we see that as time progresses, the magnetic field becomes increasingly twisted around the separator near the vertical midplane. Field lines which start within a particular topological domain at $t=0$ may (at later times) pass through several such domains before aligning with a particular null-spine.

This evidence provides an explanation for the behaviour seen in Fig.~\ref{subfig:Cpangle200eV}. By seeding the particles with a wide range of initial pitch angles, we effectively prescribe the initial amount of parallel velocity for each particle. Until the electric field completely dominates the particle behaviour (which may take several time steps), particles will be at different locations along the same field line as the magnetic field connectivity begins to change (as reconnection takes place), just when the electric field begins to fully control the particle motion. This suggests that the particles will be accelerated along field lines whose connectivity is changing 
and the exact field line along which a particle will travel will therefore depend on the position and time at which it begins to accelerate, which are pre-determined by the initial pitch angle.

A similar effect is observed for protons. However, the choice of spine along which a proton leaves no longer conforms to such a clear pattern, instead appearing to be almost random. This is likely due to the protons spending more time in the reconnection region than electrons, where each proton would experience a local environment which is changing much more rapidly.

Despite minor fluctuations of the peak kinetic energy due to this effect (of approximately $1\rm{keV}$), all electrons continue to be accelerated to $0.62\rm{MeV}$ energies ($0.07\%$ of maximum possible energy gain for the experiment). Protons are also accelerated to near identical energies, irrespective of pitch angle at position C. Therefore, neither the initial pitch angle nor kinetic energy have any significant impact on the peak kinetic energy achieved by particle in this experiment; the dominant control parameter is the initial particle position [for a given field setup, where $a$, $l$ and $b_1$ (which control the size and strength of the reconnection event taking place over a timescale $\tau$) are all fixed].

Figure~\ref{fig:twist} also provides an explanation for the oscillatory behaviour seen earlier in the electric field encountered by proton B in Fig.~\ref{subfig:PKEvEpar}. Due to their larger mass, protons are more slowly accelerated compared to electrons, and therefore remain in the reconnection region longer. In so doing, the field lines along which they travel continue to reconnect and change connectivity close to where the proton is currently located, which affects not only the proton's trajectory, but also its local environment, i.e. the electric field it encounters.


\section{Behaviour of large distributions of particles}\label{sec:globalbehaviour}

Having studied the behaviour of several different examples of particle motion within our model, we are now well-equipped to better understand the response of a large number of particles to the system of electric and magnetic fields in the vicinity of a reconnecting separator. 

Retaining the majority of parameters in our investigation from the previous section, we will look to vary the radial parameter $a$, in order to assess the limitations of the model and its viability as a realistic source of particle acceleration in the solar corona. To that end, we distribute 1280 particles, each with an initial pitch angle of $45\degr$, in a grid centred on the separator; the grid consists of five planes at different vertical heights ($z=-20,-10,0,10,20\,\rm{Mm}$), while in each plane we distribute particles in an equally spaced $16\times16$ grid array, ranging from $-0.3\rm{km}\rightarrow0.3\rm{km}$. We begin the global phase of the investigation with $a=10^{-6}z_0$; the results of a survey of electrons can be seen in Fig.~\ref{fig:a1e-6}, while the equivalent proton results can be found in Fig.~\ref{fig:a1e-6p}. In both cases, particles are given initial kinetic energies of either $2\rm{eV}$ (Figs.~\ref{subfig:a1e-6tracks}-\ref{subfig:a1e-6inipos} and Figs.~\ref{subfig:a1e-6ptracks}-\ref{subfig:a1e-6pinipos}) or $200\rm{eV}$ (Figs.~\ref{subfig:a1e-6tracks2}-\ref{subfig:a1e-6inipos2} and Figs.~\ref{subfig:a1e-6ptracks2}-\ref{subfig:a1e-6pinipos2}).

A wide range of particle energy gains are recovered in Figs.~\ref{fig:a1e-6} and~\ref{fig:a1e-6p}; both figures display variations from a minimum value near numerical accuracy ($10^{-12}\rm{eV}$, i.e. virtually no energy gain), up to a maximum energy gain of $1.4\rm{MeV}$ ($1.6\%$ of maximum possible energy gain). As seen earlier in Sec.~\ref{subsec:inipos}, the initial position of the particle plays a key role in determining not only the particle trajectory, but also the peak kinetic energy achieved. Two distinct types of behaviour are readily apparent within Fig.~\ref{fig:a1e-6}. Electron orbits which start closer to the separator always achieve larger peak kinetic energies (shown in Fig.~\ref{subfig:a1e-6inipos}); these electron orbits are strongly accelerated by the electric field up along the separator and out parallel to the spine of the upper null. Electron orbits which start further from the null experience far smaller energy gains, and typically bounce between magnetic mirror points along the separator for the entire duration of the simulation. It is also noteworthy that electrons (and protons) only escape the computational domain along field lines near the spines of one of the nulls. This is similar to the findings of \citet{paper:DallaBrowning2006} who studied the acceleration of particles at a single 3D magnetic null point in a model of spine reconnection and found that particles in their configuration escaped along the spine of the null point. In our case the particles escape along field lines that lie close to, but not exactly on, the spines of the nulls and the magnetic null points are far away from the actual reconnection site. This suggests that in our model the fact that particles preferably "escape" close to the spine of one the magnetic nulls is an effect of the geometry of magnetic field lines.

A similar picture emerges in Figs.~\ref{subfig:a1e-6tracks2}-\ref{subfig:a1e-6inipos2}, for electrons with initial kinetic energies of $200\rm{eV}$. Increasing the initial kinetic energy (as discussed in Sec.~\ref{subsec:inikeandtheta}) also increases the likelihood that an electron may mirror within the simulation domain, particularly for electrons which are not strongly accelerated by the electric field. In Fig.~\ref{subfig:a1e-6tracks2} a large percentage of the electrons in the simulation now encounter mirror points, parallel to the spines of/close to both nulls. Even electrons which achieve $\rm{keV}$ energies can become trapped in the simulation domain and mirror between the spines of both nulls. Despite $\rm{MeV}$ electrons being currently able to "escape", this is only due to our choice of computational domain (see Sect. \ref{sec:conclusions}). We also note that increasing the initial kinetic energy from $2\rightarrow200\rm{eV}$ has no effect on the peak kinetic energy gain achieved. The reason for this is that both initial energies are extremely small in comparison to the peak energies reached by the particles.

Turning to the proton behaviour observed in Fig.~\ref{fig:a1e-6p}, again we note that the protons are accelerated in the opposite direction to the electrons, due to the charge difference between each species. Several other differences between the proton results in Fig.~\ref{fig:a1e-6p} and electron results in Fig.~\ref{fig:a1e-6} are readily apparent. For the current experimental conditions, there are no visible examples of proton mirror points; none of the proton orbits in Figs.~\ref{subfig:a1e-6ptracks} and~\ref{subfig:a1e-6ptracks2} ever approach the upper null. The spread of the proton trajectories seen in Figs.~\ref{subfig:a1e-6ptracks} and~\ref{subfig:a1e-6ptracks2} is much narrower, concentrating along the separator and the spines of the lower null. 

The distribution of energy in each $x,y$ plane also changes when switching from electrons to protons. The electrons in Figs.~\ref{subfig:a1e-6inipos} and~\ref{subfig:a1e-6inipos2} show a gradual change in energy, from small gains at large values of $x,y$ to large gains close to the separator. Each $x,y$ plane contains a broad region of electrons which experience keV-MeV energy gains; each is symmetric but not radial and extends along two of the fan planes. Energy gains also vary with $z$; more electrons are found at keV-MeV energies whose starting positions are below the vertical midplane. By contrast, the proton distributions in Figs.~\ref{subfig:a1e-6pinipos} and~\ref{subfig:a1e-6pinipos2} show clear radial distributions of energy gains, where there is a sharp transition from small to large energy gains. Each region of keV-MeV energy gain is much narrower than in the electron case, and is more likely to be found above the vertical midplane. These effects result from the difference in proton and electron mass. Electrons are lighter and require much less force to be accelerated, thus effectively broadening the range of influence of the electric field (which would be otherwise ignored by protons with identical initial positions). It is also worth noting that the (roughly circular) distribution of proton energies within the planes of Fig.~\ref{subfig:a1e-6pinipos}-\ref{subfig:a1e-6pinipos2} is caused by the Gaussian decay component of Eqs.~(\ref{eq:Br})-(\ref{eq:E}). This pattern, while clear for protons, is less visible in the electron distributions, again due to the extended range of influence of the electric field, and the complex geometry of the magnetic field further from the separator.

As mentioned earlier, we will now investigate what effect the value of $a$ has on the global behaviour of the simulations. Due to current-sheet fragmentation (for example), one might perceive the value of $a=10^{-6}z_0$ to be an upper bound on the current sheet width. For our next experiment, we reduced the value of $a$ by a factor of ten, for an identical grid of electrons as discussed above. The results of this experiment can be seen in Fig.~\ref{fig:a1e-7samepos}.

By only reducing the value of $a$, the initial positions now lie outside the range of influence of the electric field, for all electrons (which, from earlier results, experience a larger range of influence of the electric field than protons). Figure~\ref{fig:a1e-7samepos} shows that every electron in this second experiment now travels upward along the separator until bouncing at a magnetic mirror point in the vicinity of the upper null; all electrons retain a peak kinetic energy equal to their initial kinetic energy ($2\rm{eV}$); identical behaviour is recovered even when the initial kinetic energy is increased (to $200\rm{eV}$). Only by placing our particles closer to the separator can we ensure that the electric field exerts some influence over the particles in question. To demonstrate this, we repeat the same experiment, with $a=10^{-7}z_0$, but reducing our particle grid spacing by a factor of 10 in each plane (i.e. particles are distributed from $-30\rm{m}\rightarrow30\rm{m}$ in $x$ and $y$); the experimental results for electrons are shown in Fig.~\ref{fig:a1e-7} and for protons in Fig.~\ref{fig:a1e-7p}.

By reducing the grid spacing and the value of $a$ together, Fig.~\ref{fig:a1e-7} shows that we continue to recover a similar distribution of energy gains in each plane as seen at $a=10^{-6}z_0$, but this time up to a maximum gain of $0.14\rm{MeV}$. The paths of electrons/protons (Fig.~\ref{subfig:a1e-7tracks}/Fig.~\ref{subfig:a1e-7ptracks}) are distributed closer to the upper/lower null, while the high energy particle distributions (Figs.~\ref{subfig:a1e-7inipos}/\ref{subfig:a1e-7pinipos}) remain similar to those observed in previous cases (see e.g. Figs.~\ref{fig:a1e-6}/\ref{fig:a1e-6p}).
\begin{figure*}[t]
 \centering
 \begin{tabular}{ccc}
  \subfloat[Electron energy spectra, $KE_{ini}=2$eV] {\label{subfig:spec_e_2eV}\resizebox{0.48\textwidth}{!}{\includegraphics{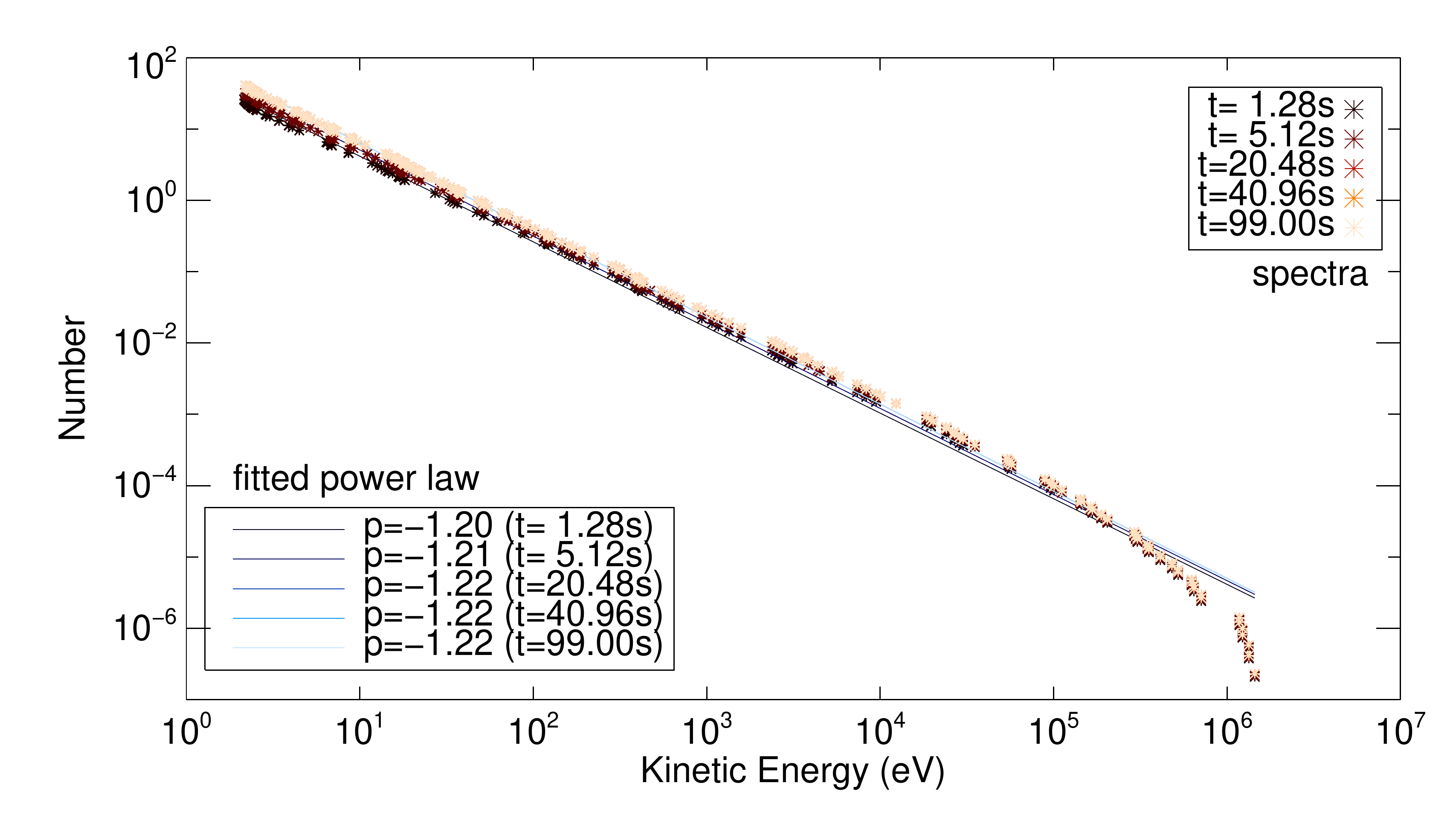}}}	&
  \subfloat[Proton energy spectra, $KE_{ini}=2$eV] {\label{subfig:spec_p_2eV}\resizebox{0.48\textwidth}{!}{\includegraphics{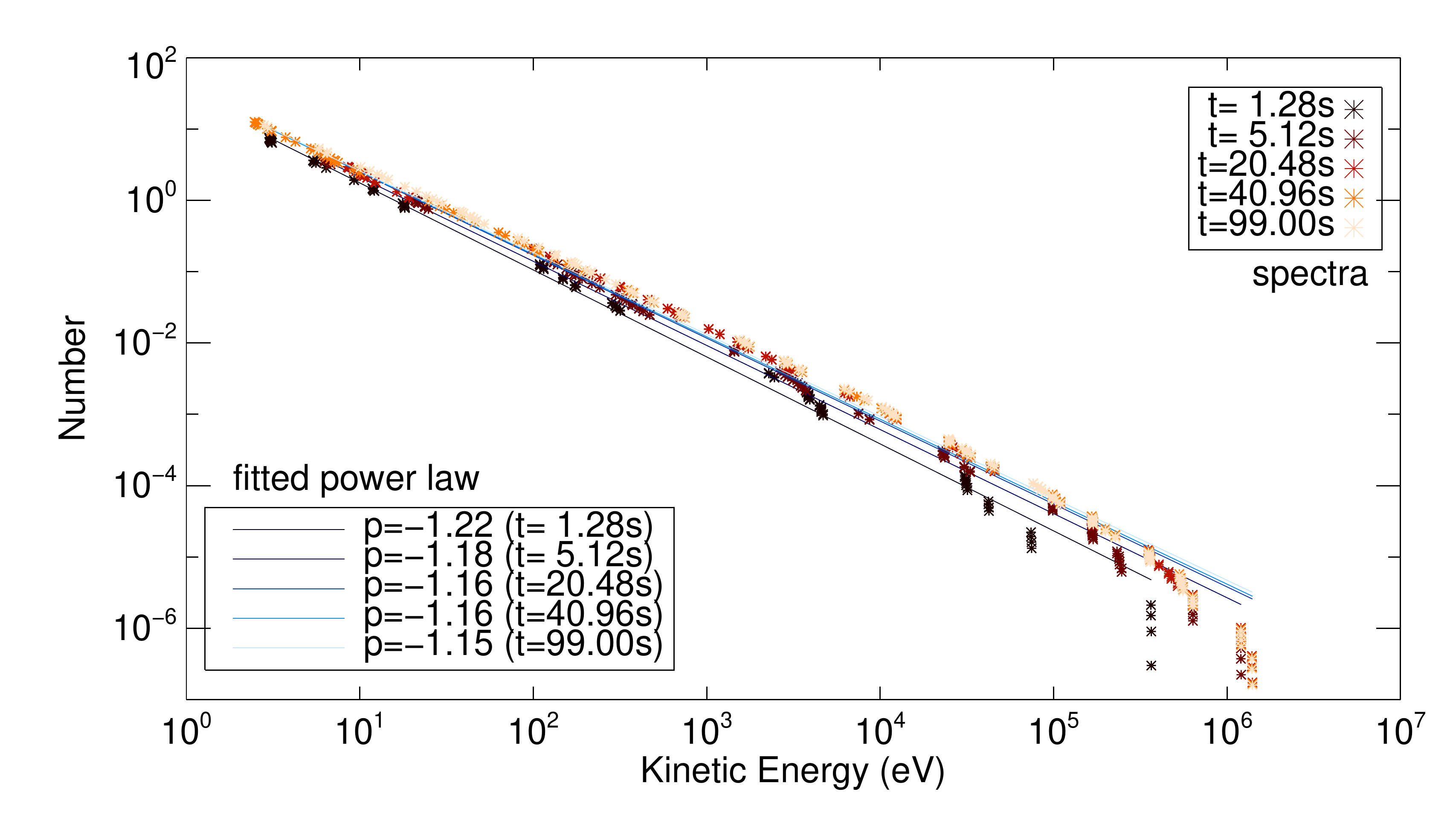}}} 		\\
  \subfloat[Electron energy spectra, $KE_{ini}=200$eV] {\label{subfig:spec_e_200eV}\resizebox{0.48\textwidth}{!}{\includegraphics{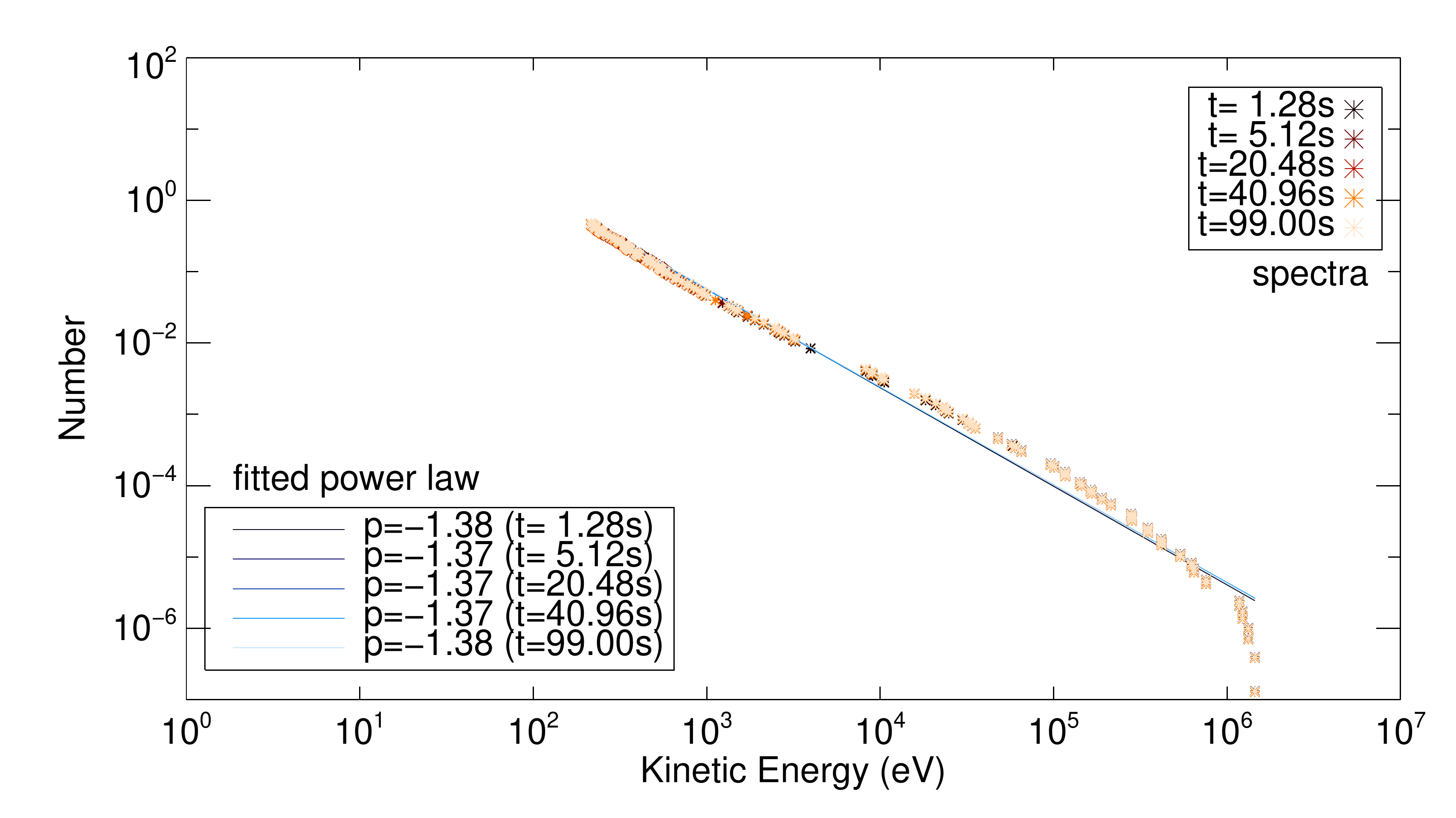}}}	&
  \subfloat[Proton energy spectra, $KE_{ini}=200$eV] {\label{subfig:spec_p_200eV}\resizebox{0.48\textwidth}{!}{\includegraphics{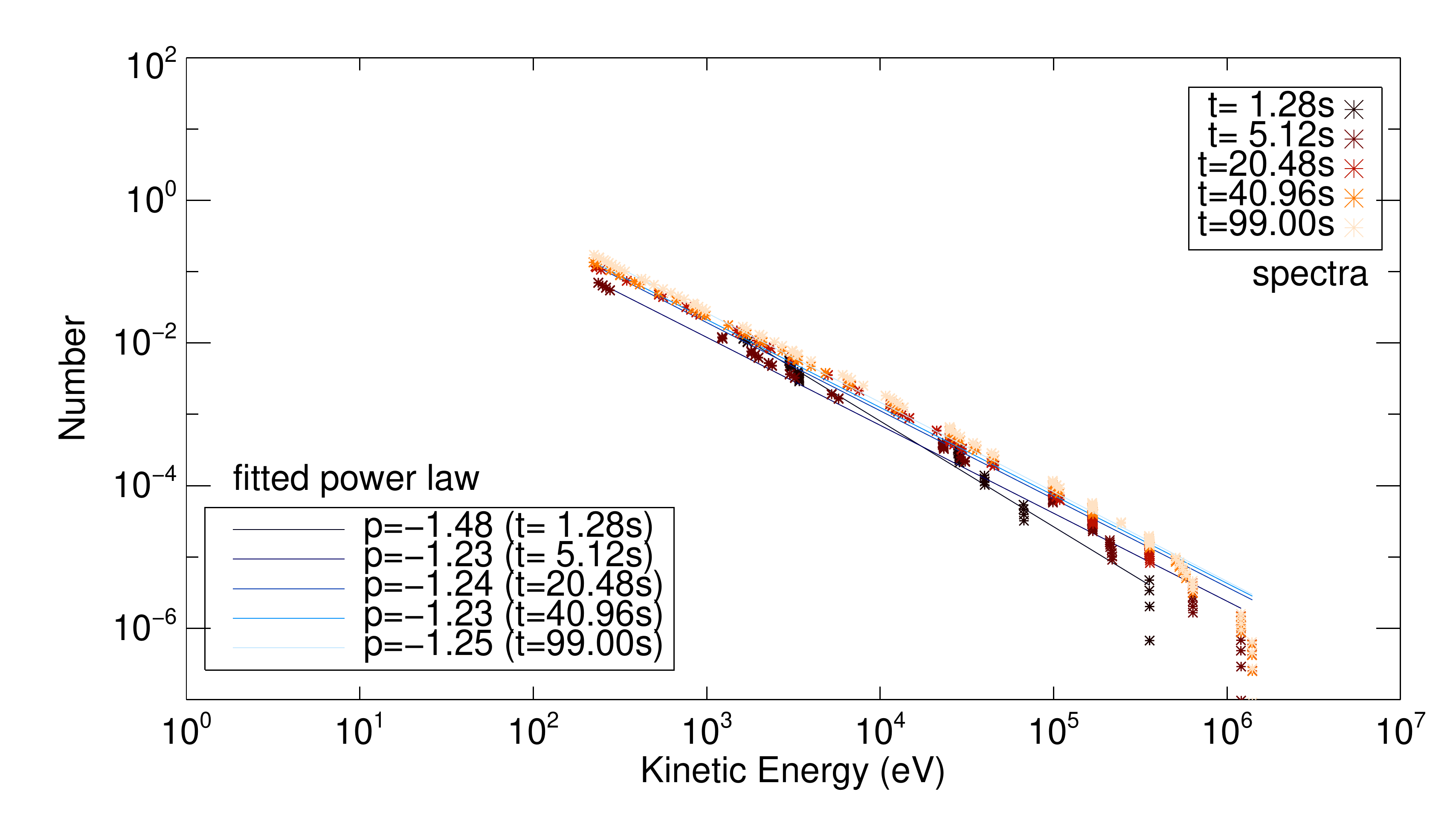}}}
 \end{tabular}
 \caption{Time evolution of energy spectra. Empirical probability distribution functions (and fitted power laws) of recorded energies at various times throughout the experiment, for \protect\subref{subfig:spec_e_2eV} Electrons all with $2\rm{eV}$ energy, \protect\subref{subfig:spec_p_2eV} Protons with $2\rm{eV}$ initial energies, \protect\subref{subfig:spec_e_200eV} Electrons with $200\rm{eV}$ energies and \protect\subref{subfig:spec_p_200eV} Protons with $200\rm{eV}$ initial energy. For key to output times for each spectra and accompanying power law, see legend.}
 \label{fig:spectra}
\end{figure*}

Many previous studies of particle acceleration present the energy spectra recorded for a given experiment. In order to stimulate discussion and further work, we too present the energy spectra of the accelerated particles previously discussed, determined at a range of times over the course of our simulations (Fig.~\ref{fig:spectra}). If a particle leaves the domain at any point, we record the particle energy on exiting the box for the remaining times. Figure~\ref{fig:spectra} shows the spectra for particles which reach energies of at least $105\%$ of their initial energy (in order to focus on the accelerated particle population alone). A power law is fitted to these spectra, using the method of maximum likelihood \citep[see e.g.][]{inbook:FeigelsonBabu2012}. The resulting probability distribution function takes the form of
\[
f_{pow}(E_k;p)=\frac{-(p+1)}{E_k^0}\left(\frac{E_k}{E_k^0}\right)^{p}, \qquad E_k\geq E_k^0,
\]
where $E_k^0$ is the minimum value of the kinetic energy, $E_k$, and $p$ is the power law index. In Figs.~\ref{fig:spectra}\subref{subfig:spec_e_2eV}-\subref{subfig:spec_p_2eV}, the initial particle energy is $2$eV and hence $E_k^0=2.1$eV; for Figs.~\ref{fig:spectra}\subref{subfig:spec_e_200eV}-\subref{subfig:spec_p_200eV}, the initial kinetic energy rises to $200$eV, hence $E_k^0=210$eV.

From Fig.~\ref{fig:spectra}, it is clear that the energy spectra recovered are relatively hard; overall values of power law parameter $p$ vary from $p\in[-1.15,-1.48]$. We note that a comparison of Fig.~\ref{subfig:spec_e_2eV} with Fig~\ref{subfig:spec_p_2eV} (or Fig.~\ref{subfig:spec_e_200eV} with Fig.~\ref{subfig:spec_p_200eV}) shows that while the electron spectra (and hence values of $p$) remain relatively fixed in time, proton spectra gradually become shallower (hence values of $p$ increase with time). This is due to the difference in mass of the particles causing electron acceleration to take place much more rapidly than proton acceleration. Many of the highly accelerated electrons leave the experiment within the first few seconds, meaning that over time the spectra remain the same. However, the protons take longer to accelerate to similar energies.


\section{Discussion}\label{sec:discussion}
We have studied both individual particle orbits (Sec.~\ref{sec:localbehaviour}) and larger sets of initial conditions for particle orbits (Sec.~\ref{sec:globalbehaviour}) to investigate the behaviour of electrons and protons in the vicinity of a reconnecting magnetic separator. From our guiding-centre and kinematic model approach presented earlier, we are able to broadly group the behaviour we found into two categories; the determining factor in this categorisation is the strength of the electric field felt by the particles.

The first of these categories concerns particles which encounter weak/negligible electric field; such particles are controlled to a large extent by their initial conditions. Particles which never fall under the influence of the electric field retain their initial kinetic energy for all time. In the magnetic field prescribed by Eq.~\ref{eq:B0} (whose field strength at large distances from the separator grows as $r^2$) \emph{all} particles initialised within our numerical box will ultimately mirror (due to the ever-increasing magnetic field strength experienced). The location of the mirror points is determined by a combination of factors; the mass of the particle, the amount of initial kinetic energy of each particle, and how this energy is split between parallel and gyro-motion (i.e. at a given pitch angle). This is illustrated by Figs.~\ref{subfig:Apangle}-\ref{subfig:Apangle200eV}. With other initial conditions unchanged, particle orbits with initial pitch angles close to $90\degr$ will mirror earlier than orbits with initial pitch angles close to $0\degr/180\degr$. Initial pitch angles in the latter range will lead to more field aligned motion initially, which would cause these particle orbits to mirror later.

The second category concerns particles which encounter strong electric field; particles in this category are rapidly accelerated along their present field line, in a direction which depends on the charge of the particle and the orientation of the local electric field. Often these particles appear to escape the ``magnetic bottle'' prescribed by our field, but these particles will ultimately mirror outside our chosen computation domain. 

Depending on the model parameters chosen, we have shown that it is possible to accelerate both electrons and protons to $10-100\rm{keV}$ within seconds (electrons) or tens-hundreds of seconds (protons) during separator reconnection. This lies well within the thick target model requirements for the interpretation of hard X-ray (HXR) emission at flaring loop footpoints in large flares \citep[see e.g.][]{paper:Brown1971,paper:Brownetal2009,review:Cargilletal2012}, in terms of energy, but does not solve the problem of the particle number fluxes required. In Sec.~\ref{sec:localbehaviour}, we demonstrate that the energies gained are effectively independent of both initial kinetic energy and pitch angle; the initial position of the particle is the primary factor in deciding its behaviour. 

We have also noted an interesting effect caused by the competition of initial pitch angle and the local electric field for electrons seeded with large initial kinetic energies; in Fig.~\ref{subfig:Cpangle200eV}, we show that electrons leave the simulation domain close to opposing magnetic spines of a particular null, grouped by pitch angle. We believe this effect to be caused by the parallel current along the separator, which is associated with a localised twisting of the magnetic field, as shown in Fig.~\ref{fig:twist}. This effect is also highlighted in Fig.~12 of \citet[][]{review:Pontin2011}, in order to demonstrate the relative complexity of such a model compared with early separator reconnection models. Protons are affected to a greater extent than the electrons by the twisting of magnetic field around the separator since they travel slower than electrons and, therefore, remain in the reconnection region for longer.

Both categories are visible in the surveys of a large number of particles with different initial conditions presented in Sec.~\ref{sec:globalbehaviour}. Protons which achieve large energy gains (in the keV-MeV scale for the current experiment parameters) are typically found in a circle centered closely on the separator, which also slightly widens as distance above the midplane increases. Outside of this, we recover significant numbers of particles which are not accelerated. For electrons this pattern is complicated by a wider range of influence of the electric field and the complicated magnetic geometry of the reconnection region. 

The fall-off in energy with radius is due to the form of the prescribed electric field (Eq.~\ref{eq:E}) which contains a Gaussian term that decays rapidly with radius. Increasing the initial kinetic energy does not significantly increase the peak kinetic energy gain, which is almost entirely determined by the electric field strength. The initial kinetic energy does affect the typical particle paths, however, allowing more particles to mirror and re-enter the reconnection region. Reducing the width parameter $a$ not only affects the radius over which particles feel the effect of the electric field, but (again noting Eq.~\ref{eq:E}) also the strength of the electric field. Thus, reducing the parameter $a$ by a factor of ten also limits the peak energy gained by any given particle by the same factor.
The parameter $a$ is also one of several parameters which directly influences the reconnection rate. In a similar calculation to that detailed in Section 5 of \citet{paper:Wilmot-SmithHornig2011}, it is possible to estimate the reconnection rate of our separator reconnection event; for the chosen parameters, this is approximately $8.85\times10^7$V. These values agree well with the reconnection rates determined from numerical experiments of separator reconnection. As an example, the reconnection rate and lengthscales used in our study fall exactly within the range of rates and scales recovered from a dynamic flux emergence experiment studied in \citet{paper:Parnelletal2010b} which yielded multiple magnetic separators, strengthening our conclusion that the electric fields and reconnection event studied here reproduce behaviour which is expected to occur within the solar atmosphere. 

It is also noteworthy that, due to our choice of model, our results are (in principle) entirely scalable; by varying the parameters within the field model, one can simply scale up/down the particle results presented here by an appropriate factor (due to the separation of scales between field and particle models). 

Finally, we turn to the energy spectra recovered by our experiments (Fig.~\ref{fig:spectra}). These spectra are well matched to power-law fits whose indices varies from $p\in[-1.15,-1.48]$. While flatter than some estimates, these indices are relatively close to those recovered by other particle acceleration models. For instance, \citet[][]{paper:Baumannetal2013} recovered a non-thermal energy tail with a power law index of $-1.78$, while \citet{paper:Stanieretal2012} record an approximate power law index of $-1.5$ for protons with energy between $10^5-10^{7.5}$eV. It should be noted thought that our power law fit is particularly well matched to our recovered spectra over many orders of magnitude; this is an essential facet of a true power law seen in nature which is not necessarily recovered in many cases in the literature which purport to show a power-law distribution. 

Observational evidence also suggests that our recovered spectra are relatively flat/hard compared to those recovered by studies of HXR fluxes during solar flares. However, examples of spectral power law indices which approach the values recovered by observations of solar flares can be found in the literature \citep[see e.g.][]{paper:Crosbyetal1993,paper:Kruckeretal2007,review:Hannahetal2011}. This raises several important issues; our relatively simple experiment is not intended to directly model a flare, merely to study how particles might respond to a separator reconnection event within the solar atmosphere. Furthermore, our initial investigation uses a uniform distribution of particle energy, pitch angle and initial position. Additional layers of complexity (including a more `realistic' set of initial conditions) may form the basis of future investigations. For this initial experiment, we felt it was important to begin with a simple picture, in order to establish the basic type(s) of particle behaviour recovered by this type of 3D reconnection.

\section{Conclusions and future work}\label{sec:conclusions}
In this investigation, we have for the first time studied the relativistic guiding centre motion of particles in a simple large-scale reconnecting 3D magnetic separator environment. Depending on the specific choice of model parameters, we have shown that separator reconnection can accelerate particles to high energies over relatively short timescales under typical solar coronal conditions. Our work highlights that weak, extended electric fields can accelerate both protons and electrons to high energies, which (for this model) primarily depend on the strength and extent of the electric field at the reconnection site. 

Accelerated particles are ejected from the reconnection site along the fan-plane field lines that run close to the spines of the two nulls which are linked by the separator; this suggests that, while we do not directly model a flare, separator reconnection processes taking place during a flare might lead to particles which impact the chromosphere, for instance, at specific localised sites. This is in contrast to previous models of particle acceleration at 3D null points in which particles may be ejected not just along the spine of the null, but also along its separatrix surface (fan plane) with no clear preferential direction, thus potentially creating a diffuse extended HXR site (again assuming that 3D null point reconnection takes place during a solar flare). 

An interesting feature of our work is that we have shown that protons may be ejected along different sets of field lines to electrons, therefore suggesting that the corresponding HXR sites might have quite different characteristics. This finding may be regarded as a three-dimensional equivalent of the finding of \citet{paper:ZharkovaGordovskyy2004} for a 2D reconnecting current sheet with guide field. We remark, however, that the equivalence is not perfect, because the particle orbits in our electromagnetic field model may eventually reach mirror points and start a bouncing motion, which is not the case in the 2D current sheet configuration.

It is important to again stress that we do not attempt to directly model a solar flare with this model. However, if one were to regard separator reconnection as playing a role in such an event, then one final conclusion our work highlights would be that the number of resulting HXR sites would be highly dependent on where the spines of the nulls actually link to. For example, it would be important to know whether these spines connect down to the chromosphere/photosphere, or extend up into interplanetary space. Furthermore, do the spines traverse large or short distances, and do they encounter regions of diverging/converging magnetic field? Due to the simplicity of the current model these questions cannot be answered, but their importance warrants further investigation.

Following on from the present investigation, several opportunities for further study present themselves. Using the present model, we intend to investigate how the role of multiple magnetic separators (which are generated much later in this experiment, after the currently studied set of high-energy particles have achieved relativistic velocities and left the numerical domain) affects the particle behaviour presented in this paper. Recent experiments have shown that multiple reconnecting magnetic separators are not uncommon \citep[see e.g.][]{paper:Haynesetal2007,paper:Parnelletal2010a,paper:Parnelletal2010b,paper:Wilmot-SmithHornig2011}, so a natural extension to this work would be to investigate their role/impact in particle acceleration.

While our use of the \citet{paper:Wilmot-SmithHornig2011} separator model has many advantages for our present work (full analytical description, highly customizable, etc), we also intend to move beyond a simple kinematic framework. Our ultimate goal is to adapt our guiding centre scheme to use input fields which are determined through the use of full 3D MHD simulations of (multiple) separator reconnection. This approach has previously been used to study particle behaviour in complex coronal structures, for example in twisted coronal loops \citep[see e.g.][]{paper:GordovskyyBrowning2011,paper:Gordovskyyetal2014} using the {\tt{Lare3d}} code \citep{paper:LareXd2001}. We therefore also see this present study as an initial proof-of-concept for later investigations, as we move towards studying particle acceleration in more complex separator reconnection scenarios.

\begin{acknowledgements}
 The authors would like to thank A.~Haynes (University of St Andrews) for assistance with 3D graphical output routines. They also gratefully acknowledge the support of the U.K. Science and Technology Facilities Council [Consolidated Grant ST/K000950/1 and a Doctoral Training Grant (SEO)]. The research leading to these results has received funding from the European Commission's Seventh Framework Programme FP7 under the grant agreement SHOCK (project number 284515) . 
\end{acknowledgements}

\bibliographystyle{aa}        
\bibliography{24366}          
\end{document}

%% file: math_def.tex
\newcommand{\beq}{ \begin{eqnarray} }
\newcommand{\eeq}{ \end{eqnarray} }

\newcommand{\xhat}{ {{\bf{\hat x}}} }
\newcommand{\yhat}{ {{\bf{\hat y}}} }
\newcommand{\zhat}{ {{\bf{\hat z}}} }

\newcommand{\curl }{ {\bf {\nabla}} \times }

\newcommand{\vpar}{ v_{\parallel} }
\newcommand{\vperp}{ v_{\perp} }
\newcommand{\Eperp}{ E_{\perp} }

\newcommand{\boldnabla}{\mbox{\boldmath$\nabla$}}

\def\grad{\boldnabla}

\def\xhat{\bf{\hat{x}}}
\def\yhat{\bf{\hat{y}}}
\def\zhat{\bf{\hat{z}}}

\def\vscl{{v_{scl}}}
\def\vsclsq{{{v_{scl}}^2}}
\def\lscl{{l_{scl}}}
\def\tscl{{t_{scl}}}
\def\bscl{{b_{scl}}}
\def\escl{{{e_{scl}}}}

\def\KEscl{{{{KE}_{scl}}}}
\def\omscl{{{\Omega_{scl}}}}

\newcommand{\Epar}{E_{\parallel}}
\newcommand{\upar}{u_{\parallel}}
\newcommand{\ue}{{{\bf{u}}_E}}


%% file: 24366.bbl
\newcommand{\noop}[1]{}
\begin{thebibliography}{65}
\expandafter\ifx\csname natexlab\endcsname\relax\def\natexlab#1{#1}\fi

\bibitem[{{Arber} {et~al.}(2001){Arber}, {Longbottom}, {Gerrard}, \&
  {Milne}}]{paper:LareXd2001}
{Arber}, T.~D., {Longbottom}, A.~W., {Gerrard}, C.~L., \& {Milne}, A.~M. 2001,
  Journal of Computational Physics, 171, 151

\bibitem[{{Arzner} \& {Vlahos}(2004)}]{paper:ArznerVlahos2004}
{Arzner}, K. \& {Vlahos}, L. 2004, \apjl, 605, L69

\bibitem[{{Arzner} \& {Vlahos}(2006)}]{paper:ArznerVlahos2006}
{Arzner}, K. \& {Vlahos}, L. 2006, \aap, 454, 957

\bibitem[{{Baumann} {et~al.}(2013){Baumann}, {Haugb{\o}lle}, \&
  {Nordlund}}]{paper:Baumannetal2013}
{Baumann}, G., {Haugb{\o}lle}, T., \& {Nordlund}, {\AA}. 2013, \apj, 771, 93

\bibitem[{{Birn} \& {Priest}(2007)}]{book:BirnPriest}
{Birn}, J. \& {Priest}, E. 2007, Reconnection of Magnetic Fields (New York:
  Cambridge University Press)

\bibitem[{{Brown}(1971)}]{paper:Brown1971}
{Brown}, J.~C. 1971, \solphys, 18, 489

\bibitem[{{Brown} {et~al.}(2009){Brown}, {Turkmani}, {Kontar}, {MacKinnon}, \&
  {Vlahos}}]{paper:Brownetal2009}
{Brown}, J.~C., {Turkmani}, R., {Kontar}, E.~P., {MacKinnon}, A.~L., \&
  {Vlahos}, L. 2009, \aap, 508, 993

\bibitem[{{Browning} \& {Vekstein}(2001)}]{paper:BrowningVekstein2001}
{Browning}, P.~K. \& {Vekstein}, G.~E. 2001, \jgr, 106, 18677

\bibitem[{{Bruhwiler} \& {Zweibel}(1992)}]{paper:BruhwilerZweibel1992}
{Bruhwiler}, D.~L. \& {Zweibel}, E.~G. 1992, \jgr, 97, 10825

\bibitem[{{Bulanov} \& {Sasorov}(1976)}]{paper:BulanovSasorov1976}
{Bulanov}, S.~V. \& {Sasorov}, P.~V. 1976, \sovast, 19, 464

\bibitem[{{Cargill} {et~al.}(2012){Cargill}, {Vlahos}, {Baumann}, {Drake}, \&
  {Nordlund}}]{review:Cargilletal2012}
{Cargill}, P.~J., {Vlahos}, L., {Baumann}, G., {Drake}, J.~F., \& {Nordlund},
  {\AA}. 2012, \ssr, 173, 223

\bibitem[{{Crosby} {et~al.}(1993){Crosby}, {Aschwanden}, \&
  {Dennis}}]{paper:Crosbyetal1993}
{Crosby}, N.~B., {Aschwanden}, M.~J., \& {Dennis}, B.~R. 1993, \solphys, 143,
  275

\bibitem[{{Dalla} \& {Browning}(2005)}]{paper:DallaBrowning2005}
{Dalla}, S. \& {Browning}, P.~K. 2005, \aap, 436, 1103

\bibitem[{{Dalla} \& {Browning}(2006)}]{paper:DallaBrowning2006}
{Dalla}, S. \& {Browning}, P.~K. 2006, \apjl, 640, L99

\bibitem[{{Dalla} \& {Browning}(2008)}]{paper:DallaBrowning2008}
{Dalla}, S. \& {Browning}, P.~K. 2008, \aap, 491, 289

\bibitem[{{Deng} {et~al.}(2009){Deng}, {Zhou}, {Li}, {Baumjohann}, {Andre},
  {Cornilleau}, {Santol{\'{\i}}k}, {Pontin}, {Reme}, {Lucek}, {Fazakerley},
  {Decreau}, {Daly}, {Nakamura}, {Tang}, {Hu}, {Pang}, {B{\"u}chner}, {Zhao},
  {Vaivads}, {Pickett}, {Ng}, {Lin}, {Fu}, {Yuan}, {Su}, \&
  {Wang}}]{paper:Dengetal2009}
{Deng}, X.~H., {Zhou}, M., {Li}, S.~Y., {et~al.} 2009, Journal of Geophysical
  Research (Space Physics), 114, 7216

\bibitem[{{Dorelli} \& {Bhattacharjee}(2008)}]{paper:DorelliBhattacharjee2008}
{Dorelli}, J.~C. \& {Bhattacharjee}, A. 2008, Physics of Plasmas, 15, 056504

\bibitem[{{Drake} {et~al.}(2006){Drake}, {Swisdak}, {Che}, \&
  {Shay}}]{paper:Drakeetal2006}
{Drake}, J.~F., {Swisdak}, M., {Che}, H., \& {Shay}, M.~A. 2006, \nat, 443, 553

\bibitem[{Feigelson \& Babu(2012)}]{inbook:FeigelsonBabu2012}
Feigelson, E. \& Babu, G. 2012, Modern Statistical Methods for Astronomy: With
  R Applications (Cambridge University Press), 90

\bibitem[{{Fletcher} {et~al.}(2011){Fletcher}, {Dennis}, {Hudson}, {Krucker},
  {Phillips}, {Veronig}, {Battaglia}, {Bone}, {Caspi}, {Chen}, {Gallagher},
  {Grigis}, {Ji}, {Liu}, {Milligan}, \& {Temmer}}]{review:Fletcheretal2011}
{Fletcher}, L., {Dennis}, B.~R., {Hudson}, H.~S., {et~al.} 2011, \ssr, 159, 19

\bibitem[{{Galsgaard} {et~al.}(2000){Galsgaard}, {Priest}, \&
  {Nordlund}}]{paper:Galgaardetal2000}
{Galsgaard}, K., {Priest}, E.~R., \& {Nordlund}, {\AA}. 2000, \solphys, 193, 1

\bibitem[{{Giovanelli}(1946)}]{paper:Giovanelli1946}
{Giovanelli}, R.~G. 1946, \nat, 158, 81

\bibitem[{{Gordovskyy} \& {Browning}(2011)}]{paper:GordovskyyBrowning2011}
{Gordovskyy}, M. \& {Browning}, P.~K. 2011, \apj, 729, 101

\bibitem[{{Gordovskyy} {et~al.}(2013){Gordovskyy}, {Browning}, {Kontar}, \&
  {Bian}}]{paper:Gordovskyyetal2013}
{Gordovskyy}, M., {Browning}, P.~K., {Kontar}, E.~P., \& {Bian}, N.~H. 2013,
  \solphys, 284, 489

\bibitem[{{Gordovskyy} {et~al.}(2014){Gordovskyy}, {Browning}, {Kontar}, \&
  {Bian}}]{paper:Gordovskyyetal2014}
{Gordovskyy}, M., {Browning}, P.~K., {Kontar}, E.~P., \& {Bian}, N.~H. 2014,
  A\&A, 561, A72

\bibitem[{{Gordovskyy} {et~al.}(2010{\natexlab{a}}){Gordovskyy}, {Browning}, \&
  {Vekstein}}]{paper:Gordovskyyetal2010a}
{Gordovskyy}, M., {Browning}, P.~K., \& {Vekstein}, G.~E. 2010{\natexlab{a}},
  \aap, 519, A21

\bibitem[{{Gordovskyy} {et~al.}(2010{\natexlab{b}}){Gordovskyy}, {Browning}, \&
  {Vekstein}}]{paper:Gordovskyyetal2010b}
{Gordovskyy}, M., {Browning}, P.~K., \& {Vekstein}, G.~E. 2010{\natexlab{b}},
  \apj, 720, 1603

\bibitem[{{Guo} {et~al.}(2010){Guo}, {B{\"u}chner}, {Otto}, {Santos}, {Marsch},
  \& {Gan}}]{paper:Guoetal2010}
{Guo}, J.-N., {B{\"u}chner}, J., {Otto}, A., {et~al.} 2010, \aap, 513, A73

\bibitem[{{Guo} {et~al.}(2013){Guo}, {Pu}, {Xiao}, {Wang}, {Fu}, {Xie}, {Zong},
  {He}, {Yao}, {Zhong}, \& {Li}}]{paper:Guoetal2013}
{Guo}, R., {Pu}, Z., {Xiao}, C., {et~al.} 2013, Journal of Geophysical Research
  (Space Physics), 118, 6116

\bibitem[{{Hannah} \& {Fletcher}(2006)}]{paper:HannahFletcher2006}
{Hannah}, I.~G. \& {Fletcher}, L. 2006, \solphys, 236, 59

\bibitem[{{Hannah} {et~al.}(2011){Hannah}, {Hudson}, {Battaglia}, {Christe},
  {Ka{\v s}parov{\'a}}, {Krucker}, {Kundu}, \&
  {Veronig}}]{review:Hannahetal2011}
{Hannah}, I.~G., {Hudson}, H.~S., {Battaglia}, M., {et~al.} 2011, \ssr, 159,
  263

\bibitem[{{Haynes} {et~al.}(2007){Haynes}, {Parnell}, {Galsgaard}, \&
  {Priest}}]{paper:Haynesetal2007}
{Haynes}, A.~L., {Parnell}, C.~E., {Galsgaard}, K., \& {Priest}, E.~R. 2007,
  Royal Society of London Proceedings Series A, 463, 1097

\bibitem[{{Hesse} \& {Schindler}(1988)}]{paper:HesseSchindler1988}
{Hesse}, M. \& {Schindler}, K. 1988, \jgr, 93, 5559

\bibitem[{{Kliem}(1994)}]{paper:Kliem1994}
{Kliem}, B. 1994, \apjs, 90, 719

\bibitem[{{Krucker} {et~al.}(2007){Krucker}, {Kontar}, {Christe}, \&
  {Lin}}]{paper:Kruckeretal2007}
{Krucker}, S., {Kontar}, E.~P., {Christe}, S., \& {Lin}, R.~P. 2007, \apjl,
  663, L109

\bibitem[{{Lau} \& {Finn}(1990)}]{paper:LauFinn1990}
{Lau}, Y.-T. \& {Finn}, J.~M. 1990, \apj, 350, 672

\bibitem[{{Litvinenko}(1996)}]{paper:Litvinenko1996}
{Litvinenko}, Y.~E. 1996, \apj, 462, 997

\bibitem[{{Longcope}(2001)}]{paper:Longcope2001}
{Longcope}, D.~W. 2001, Physics of Plasmas, 8, 5277

\bibitem[{{Longcope} \& {Cowley}(1996)}]{paper:LongcopeCowley1996}
{Longcope}, D.~W. \& {Cowley}, S.~C. 1996, Physics of Plasmas, 3, 2885

\bibitem[{{Longcope} {et~al.}(2005){Longcope}, {McKenzie}, {Cirtain}, \&
  {Scott}}]{paper:Longcopeetal2005}
{Longcope}, D.~W., {McKenzie}, D.~E., {Cirtain}, J., \& {Scott}, J. 2005, \apj,
  630, 596

\bibitem[{{Metcalf} {et~al.}(2003){Metcalf}, {Alexander}, {Hudson}, \&
  {Longcope}}]{paper:Metcalfetal2003}
{Metcalf}, T.~R., {Alexander}, D., {Hudson}, H.~S., \& {Longcope}, D.~W. 2003,
  \apj, 595, 483

\bibitem[{{Neukirch} {et~al.}(2007){Neukirch}, {Giuliani}, \&
  {Wood}}]{paper:Neukirchetal2007}
{Neukirch}, T., {Giuliani}, P., \& {Wood}, P.~D. 2007, in Reconnection of
  Magnetic Fields, ed. J.~Birn \& E.~Priest (Cambridge University Press),
  281--291

\bibitem[{Northrop(1963)}]{book:Northrop1963}
Northrop, T. 1963, The adiabatic motion of charged particles, Interscience
  tracts on physics and astronomy (Interscience Publishers)

\bibitem[{{Oskoui} \& {Neukirch}(2014)}]{paper:OskouiNeukirch2014}
{Oskoui}, S.~E. \& {Neukirch}, T. 2014, \aap, 567, A131

\bibitem[{{Parnell}(2007)}]{paper:Parnell2007}
{Parnell}, C.~E. 2007, \solphys, 242, 21

\bibitem[{{Parnell} {et~al.}(2008){Parnell}, {Haynes}, \&
  {Galsgaard}}]{paper:Parnelletal2008}
{Parnell}, C.~E., {Haynes}, A.~L., \& {Galsgaard}, K. 2008, \apj, 675, 1656

\bibitem[{{Parnell} {et~al.}(2010{\natexlab{a}}){Parnell}, {Haynes}, \&
  {Galsgaard}}]{paper:Parnelletal2010a}
{Parnell}, C.~E., {Haynes}, A.~L., \& {Galsgaard}, K. 2010{\natexlab{a}},
  Journal of Geophysical Research (Space Physics), 115, 2102

\bibitem[{{Parnell} {et~al.}(2010{\natexlab{b}}){Parnell}, {Maclean}, \&
  {Haynes}}]{paper:Parnelletal2010b}
{Parnell}, C.~E., {Maclean}, R.~C., \& {Haynes}, A.~L. 2010{\natexlab{b}},
  \apjl, 725, L214

\bibitem[{{Pontin}(2011)}]{review:Pontin2011}
{Pontin}, D.~I. 2011, Advances in Space Research, 47, 1508

\bibitem[{{Pontin} \& {Craig}(2006)}]{paper:PontinCraig2006}
{Pontin}, D.~I. \& {Craig}, I.~J.~D. 2006, \apj, 642, 568

\bibitem[{{Priest} \& {Forbes}(2000)}]{book:PriestForbes}
{Priest}, E.~R. \& {Forbes}, T. 2000, Magnetic Reconnection: MHD Theory and
  Applications (Cambridge University Press)

\bibitem[{{Schindler} {et~al.}(1988){Schindler}, {Hesse}, \&
  {Birn}}]{paper:Schindleretal1988}
{Schindler}, K., {Hesse}, M., \& {Birn}, J. 1988, \jgr, 93, 5547

\bibitem[{{Schindler} {et~al.}(1991){Schindler}, {Hesse}, \&
  {Birn}}]{paper:Schindleretal1991}
{Schindler}, K., {Hesse}, M., \& {Birn}, J. 1991, \apj, 380, 293

\bibitem[{{Sonnerup}(1979)}]{paper:Sonnerup1979}
{Sonnerup}, B.~U.~{\"O}. 1979, in Space Plasma Physics: The Study of
  Solar-System Plasmas. Volume 2, 879

\bibitem[{{Stanier} {et~al.}(2012){Stanier}, {Browning}, \&
  {Dalla}}]{paper:Stanieretal2012}
{Stanier}, A., {Browning}, P., \& {Dalla}, S. 2012, \aap, 542, A47

\bibitem[{{Stevenson} {et~al.}(2014){Stevenson}, {Parnell}, {Priest}, \&
  {Haynes}}]{subm:Stevensonetal2014}
{Stevenson}, J.~E.~H., {Parnell}, C.~E., {Priest}, E.~R., \& {Haynes}, A.~L.
  2014, \aap, submitted for publication

\bibitem[{{Turkmani} {et~al.}(2006){Turkmani}, {Cargill}, {Galsgaard},
  {Vlahos}, \& {Isliker}}]{paper:Turkmanietal2006}
{Turkmani}, R., {Cargill}, P.~J., {Galsgaard}, K., {Vlahos}, L., \& {Isliker},
  H. 2006, \aap, 449, 749

\bibitem[{{Turkmani} {et~al.}(2005){Turkmani}, {Vlahos}, {Galsgaard},
  {Cargill}, \& {Isliker}}]{paper:Turkmanietal2005}
{Turkmani}, R., {Vlahos}, L., {Galsgaard}, K., {Cargill}, P.~J., \& {Isliker},
  H. 2005, \apjl, 620, L59

\bibitem[{{Vandervoort}(1960)}]{paper:Vandervoort1960}
{Vandervoort}, P.~O. 1960, Annals of Physics, 10, 401

\bibitem[{{Vlahos} {et~al.}(2004){Vlahos}, {Isliker}, \&
  {Lepreti}}]{paper:Vlahosetal2004}
{Vlahos}, L., {Isliker}, H., \& {Lepreti}, F. 2004, \apj, 608, 540

\bibitem[{{Wilmot-Smith} \& {Hornig}(2011)}]{paper:Wilmot-SmithHornig2011}
{Wilmot-Smith}, A.~L. \& {Hornig}, G. 2011, \apj, 740, 89

\bibitem[{{Wood} \& {Neukirch}(2005)}]{paper:WoodNeukirch2005}
{Wood}, P. \& {Neukirch}, T. 2005, \solphys, 226, 73

\bibitem[{{Xiao} {et~al.}(2007){Xiao}, {Wang}, {Pu}, {Ma}, {Zhao}, {Zhou},
  {Wang}, {Kivelson}, {Fu}, {Liu}, {Zong}, {Dunlop}, {Glassmeier}, {Lucek},
  {Reme}, {Dandouras}, \& {Escoubet}}]{paper:Xiaoetal2007}
{Xiao}, C.~J., {Wang}, X.~G., {Pu}, Z.~Y., {et~al.} 2007, Nature Physics, 3,
  609

\bibitem[{{Zharkova} \& {Gordovskyy}(2004)}]{paper:ZharkovaGordovskyy2004}
{Zharkova}, V.~V. \& {Gordovskyy}, M. 2004, \apj, 604, 884

\bibitem[{{Zharkova} \& {Gordovskyy}(2005)}]{paper:ZharkovaGordovskyy2005}
{Zharkova}, V.~V. \& {Gordovskyy}, M. 2005, \mnras, 356, 1107

\end{thebibliography}
